\documentclass[12pt]{article}
\title{
Macroscopic and Microscopic Thermal Equilibrium}
\author{
Sheldon Goldstein\footnote{Department of Mathematics,
     Rutgers University, Hill Center, 
     110 Frelinghuysen Road, Piscataway, NJ 08854-8019, USA.
     E-mail: oldstein@math.rutgers.edu},
David A. Huse\footnote{Department of Physics, Princeton University, 
    Jadwin Hall, Washington Road, Princeton, NJ  08544-0708, USA and
    Institute for Advanced Study, Princeton, NJ 08540, USA.
    E-mail: huse@princeton.edu},\\
Joel L. Lebowitz\footnote{Departments of Mathematics and Physics,
     Rutgers University, Hill Center, 
     110 Frelinghuysen Road, Piscataway, NJ 08854-8019, USA.
     E-mail: lebowitz@math.rutgers.edu}, and 
Roderich Tumulka\footnote{Department of Mathematics,
     Rutgers University, Hill Center, 
     110 Frelinghuysen Road, Piscataway, NJ 08854-8019, USA.
     E-mail: tumulka@math.rutgers.edu}
}
\date{January 12, 2017}

\usepackage{amsthm,url,amssymb,amsmath,amsfonts,mathrsfs}
\usepackage[usenames,dvipsnames]{color}
\usepackage{graphicx}

\newcounter{remarks}
\newcommand{\ket}[1]{\vert#1\rangle}
\newcommand{\bra}[1]{\langle#1\vert}
\newcommand{\pr}[1]{\ket{#1}\bra{#1}}
\DeclareMathOperator{\tr}{tr}
\DeclareMathOperator{\Var}{Var}
\DeclareMathOperator{\vol}{vol}
\newcommand{\MATE}{\mathrm{MATE}}
\newcommand{\TMATE}{\mathrm{TMATE}}
\newcommand{\MITE}{\mathrm{MITE}}
\newcommand{\CCC}{\mathbb{C}}
\newcommand{\RRR}{\mathbb{R}}
\newcommand{\EEE}{\mathbb{E}}
\newcommand{\SSS}{\mathbb{S}}
\newcommand{\ZZZ}{\mathbb{Z}}

\newcommand{\scp}[2]{\langle #1| #2 \rangle}
\newcommand{\Hilbert}{\mathscr{H}}
\newcommand{\be}{\begin{equation}}
\newcommand{\ee}{\end{equation}}
\newcommand{\mc}{\mathrm{mc}}
\newcommand{\know}{\mathrm{know}}
\newcommand{\eq}{\mathrm{eq}}
\newcommand{\R}{\mathscr{R}}
\newcommand{\sA}{\mathscr{A}}
\newcommand{\sS}{\mathscr{S}}

\begin{document}
\maketitle
\begin{abstract}
We study the nature of and approach to thermal equilibrium in isolated quantum systems. An individual isolated macroscopic quantum system in a pure or mixed state is regarded as being in thermal equilibrium if all macroscopic observables assume rather sharply the values obtained from  thermodynamics. Of such a system (or state) we say that it is in macroscopic thermal equilibrium (MATE). A stronger requirement than MATE is that even microscopic observables (i.e., ones referring to a small subsystem) have a probability distribution in agreement with that obtained from the micro-canonical, or equivalently the canonical, ensemble for the whole system. Of such a system we say that it is in microscopic thermal equilibrium (MITE). The distinction between MITE and MATE is particularly relevant for systems with many-body localization (MBL) for which the energy eigenfuctions fail to be in MITE while necessarily most of them, but not all, are in MATE. However, if we consider superpositions of energy eigenfunctions (i.e., typical wave functions $\psi$) in an energy shell, then for generic macroscopic systems, including those with MBL, most $\psi$ are in both MATE and MITE. We explore here the properties of MATE and MITE and compare the two notions, thereby elaborating on ideas introduced in \cite{GHLT15a}.

\medskip

Key words: many-body localization; quantum statistical mechanics; canonical typicality; thermal equilibrium subspace; macro-observables; thermalization.
\end{abstract}

\section{Introduction}
\label{sec:intro}

The notion of a thermal equilibrium state of a macroscopic system (say, one with $N>10^{20}$ degrees of freedom) is basic to thermodynamics. Its existence is Postulate 1 in the Tisza--Callen formulation of thermodynamics \cite{Ca}. Informally, one can use Onsager's description:
\begin{quotation}
 These ``thermodynamic" states are typically defined as states of
 ``equilibrium" under specified restraints
 in composition, energy, and external boundary conditions, in that no
 spontaneous change can
 occur in the system as long as the constraints remain fixed.
 \cite{Ons} (quotation marks in original)
\end{quotation}
One would, of course, also like to have a microscopic description of what it means for a system to be in thermal equilibrium in terms of the micro-state considered in statistical mechanics, i.e., in terms of the phase space $\Gamma$ of a classical system or the Hilbert space $\Hilbert$ of a quantum system.

When speaking of thermal equilibrium, one often refers to a \emph{thermodynamic ensemble}, which corresponds classically to a probability distribution over $\Gamma$ and quantum-mechanically to a density operator on $\Hilbert$. For example, the canonical ensemble at inverse temperature $\beta$ has, classically, the density function
\be\label{rhobetadef}
\rho^{(\beta)}(X) = \frac{1}{Z} e^{-\beta H(X)}
\ee
for any $X\in \Gamma$, with normalizing constant $Z$ and Hamiltonian function $H:\Gamma\to\RRR$. 
In quantum mechanics, the canonical ensemble corresponds to the density operator
\be\label{hatrhobetadef}
\hat\rho^{(\beta)} = \frac{1}{Z} e^{-\beta \hat H}
\ee
with a different normalizing constant $Z$ and Hamiltonian operator $\hat H$ on $\Hilbert$. Likewise, the micro-canonical ensemble is, classically, the uniform density $\rho^{\mc}$ over a micro-canonical energy shell
\be\label{Gammamc}
\Gamma_{\mc} = \Bigl\{X\in\Gamma: E-\Delta E< H(X)\leq  E\Bigr\}
\ee
whose width $\Delta E$ represents the macroscopic resolution of energy. In quantum mechanics, the micro-canonical ensemble corresponds to the density operator
\be\label{hatrhomcdef}
\hat\rho^{\mc}= \frac{1}{\dim \Hilbert_{\mc}} \hat{P}_{\mc}\,,
\ee
where $\Hilbert_{\mc}$, also called the micro-canonical energy shell, is the subspace of $\Hilbert$ spanned by the eigenvectors of $\hat{H}$ with eigenvalue between $E-\Delta E$ and $E$, and $\hat{P}_{\mc}$ is the projection to $\Hilbert_{\mc}$.

However, such an ensemble does not 
answer the need for a definition of ``thermal equilibrium,'' as one often wants to consider an individual closed, macroscopic system in thermal equilibrium. For example, we want to know whether \emph{this particular} thermos bottle of coffee is in thermal equilibrium. Put differently, we assume an ``individualist'' attitude, as opposed to the ``ensemblist'' attitude \cite{GLTZ10}. An individual system corresponds classically to a point in phase space, rather than to a distribution over phase space. Also in quantum mechanics, one often wants to regard a system in a pure state $\ket{\psi}$ as being in thermal equilibrium, while its density matrix $\hat\rho=\pr{\psi}$ is far away from the $\hat\rho^{(\beta)}$ of \eqref{hatrhobetadef} and the $\hat\rho^{\mc}$ of \eqref{hatrhomcdef}. This is certainly possible; in fact, it has been an active field of research for a number of years now to study how a closed quantum system in a pure state can display thermal equilibrium behavior; see, e.g., \cite{GMM04,PSW06,GLTZ06,Sug07,Rei07,Rei08,RDO08,LPSW08,GLMTZ09b,Rei10,RS12,Rei15,GE15}, after some pioneering work even earlier \cite{vN29,JS,Deu91,Sre94}. In Section~\ref{sec:pure} we elaborate on the reasons for considering systems in pure states. 

In this paper, which elaborates on ideas introduced in \cite{GHLT15a}, we explain how the idea of thermal equilibrium of a system in a pure state can be defined for a macroscopic quantum system. 
Of particular interest in this context are systems featuring \emph{many-body Anderson localization (MBL)} \cite{And58,BAA06b,OH07}. These are quantum systems whose Hamiltonian $\hat H$ has eigenfunctions that are in a certain sense spatially localized, which can be an obstacle to reaching thermal equilibrium (in whatever sense). 

A natural definition of thermal equilibrium is to say that a system with (pure or mixed) density matrix $\hat\rho$
is in thermal equilibrium when all macro observables assume rather sharp values in $\hat\rho$ that agree with their thermodynamic equilibrium values; we call this notion \emph{macroscopic thermal equilibrium (MATE)}. As we discuss below, the pure states $\psi$ in MATE in a given micro-canonical energy shell are all close to a certain subspace of Hilbert space, the thermal equilibrium subspace $\Hilbert_{\eq}$ for this energy shell. For many systems including those with MBL, $\Hilbert_{\eq}$ has the overwhelming majority of dimensions in the energy shell, and most pure states in the energy shell are in MATE, as are most mixed states. Here and throughout this paper, ``most'' means ``the overwhelming majority of'' (or ``all except a small set'') relative to the relevant uniform distribution; for  example, ``most pure states in the energy shell'' means the overwhelming majority relative to the uniform distribution on the unit sphere in $\Hilbert_{\mc}$ (see Remark~\ref{rem:mostMATE} in Section~\ref{sec:MATE1}).

Now, for generic macroscopic systems, with or without MBL, most $\psi$'s have a stronger property: That micro observables (i.e., any observable referring to a small subsystem $S$) have a probability distribution in $\psi$ that coincides with their thermal probability distribution; we say that a system with such a $\psi$ (or, in fact, such a $\hat\rho$) is in \emph{microscopic thermal equilibrium (MITE)}. This property is a sign of a high degree of entanglement in $\psi$ between $S$ and its complement.

A dynamical aspect of our theme is the \emph{approach to thermal equilibrium}, by which we mean (for either MATE or MITE) that a system starting out away from thermal equilibrium sooner or later reaches thermal equilibrium and spends most of the time in the long run in thermal equilibrium. This behavior is connected to the \emph{eigenstate thermalization hypothesis} (ETH), which asserts that the energy eigenfunctions are in thermal equilibrium, and which therefore can be considered in two variants, as MATE-ETH or MITE-ETH. If \emph{all} energy eigenstates are in MATE, then it can be shown \cite{GLMTZ09b} that \emph{all} pure states approach MATE; for MITE, the situation is a bit more complicated, as discussed in Section~\ref{sec:overview}. For MBL systems, some pure states fail to approach either MATE or MITE, which is related to the failure of MATE for some eigenstates and MITE for all eigenstates for such systems (again, see Section~\ref{sec:overview}).

In the remainder of this paper we explore the two notions, MATE and MITE, their properties and relations to MBL. The remaining sections are organized as follows. 
In Section~\ref{sec:pure}, we describe our motivation for considering an individual system, possibly even in a pure state. 
In Section~\ref{sec:classical}, we take a look at the classical situation of thermal equilibrium. 
In Section~\ref{sec:quantum}, we give a detailed description of the concepts of MATE and MITE. 
In Section~\ref{sec:approach} we focus on the dynamical approach to MATE or MITE and the eigenstate thermalization hypothesis (ETH). 
In Section~\ref{sec:mbl}, we illustrate MATE and MITE for specific simple MBL systems. 
In Section~\ref{sec:overview}, we explore further aspects of MATE and MITE. 
In Section~\ref{sec:exceptional}, we address cases in which no dominant macro-state exists.
In Section~\ref{sec:otherdefs}, we review a couple of other proposed definitions of thermal equilibrium. We conclude in Section~\ref{sec:conclusions}.

\section{Why Include Pure States?}
\label{sec:pure}

Readers familiar and comfortable with the individualist attitude may want to skip this section.

\subsection{Classical Mechanics}

In the ensemblist attitude, one would say that thermal equilibrium occurs when, for a classical system, the probability density is close to that of a suitable thermodynamic ensemble---say, to $\rho^{(\beta)}$  or $\rho^{\mc}$. Thus, thermal equilibrium would seem to require a ``mixed state'' (i.e., a probability distribution over phase space), and not a ``pure state'' (i.e., a point in phase space). So why do we insist on considering pure states?

The reason is that an \emph{individual system} has a unique phase point $X$ (a pure state), and it seems meaningful and necessary to talk about whether this system is in thermal equilibrium. For example, we can talk about this particular thermos bottle of coffee, how the energy is spatially distributed in it, in particular whether the local temperature is constant throughout the coffee. 

To be sure, \emph{our knowledge} of the system can be represented by some probability density function $\rho_{\know}$ over the phase space, and since our knowledge is usually very limited, as we do not know the exact position and momentum of every molecule in this bottle of coffee, $\rho_{\know}$ is usually \emph{very} spread-out (not at all a pure state). However, when we ask whether the coffee in this particular bottle is in thermal equilibrium, we are not asking whether $\rho_{\know}$ is close to $\rho^{(\beta)}$ or $\rho^{\mc}$; instead, we are asking about how the energy is spatially distributed, and whether the local temperature is constant. We are asking about properties of the phase point $X$, not of our knowledge $\rho_{\know}$ (a point made particularly in \cite{LM03}). In fact, if we do not have the relevant knowledge about $X$, if we do not know the spatial distribution of energy in this particular bottle, we have to answer that we do not know whether the content of the bottle is in thermal equilibrium, and we need to make measurements on the system to find out whether it is in thermal equilibrium. We do not want to say that the system is not in thermal equilibrium just because we do not know its phase point---or because we do.

So, we say that a phase point $X$ is in thermal equilibrium if it has all the properties of thermal equilibrium, such as a uniform spatial distribution of energy over the volume of the bottle (see Section~\ref{sec:classical} below for more detail). By $\Gamma_{\eq}$ we denote the set of those $X$. Should our knowledge correspond to $\rho_{\know}=\rho^{\mc}$, then we are $>99.99\%$ confident that $X$ is in thermal equilibrium, as $\Gamma_{\eq}$ has most of the phase space volume of $\Gamma_{\mc}$.

\subsection{Quantum Mechanics}

In quantum mechanics, the situation is a bit more complicated and richer than in the classical case. That is mainly because a mixed state, i.e., a density matrix $\hat\rho$ on $\Hilbert$, can arise in two ways: either as representing our lack of knowledge (analogously to probability distributions in the classical case), or as a consequence of entanglement, i.e., as a reduced density matrix obtained by tracing out another system with which our system is entangled. For that reason, we do not insist that the system be in a pure state, but we insist that a system in a pure state can be in thermal equilibrium!

As in the classical case, we regard the experimenter's lack of knowledge as irrelevant to the question of whether the system is in thermal equilibrium. This attitude already suggests using a definition of thermal equilibrium that allows also systems in pure states to be in thermal equilibrium: Since classically a single $X$ could be in thermal equilibrium, why not a single $\psi$? Likewise, since knowing $X$ did not matter, why would knowing $\psi$ matter? As in the classical case, when we ask whether a system is in thermal equilibrium, we do not ask a question about our limited knowledge but one about the factual state of affairs. For that reason, we admit the possibility that the system may have a pure state $\psi$ that we do not know. 

Moreover, if by thermal equilibrium we mean that (e.g.)\ energy is uniformly distributed (within suitable tolerances) over the volume, then that can very well be the case also for a pure quantum state $\psi$. 
(For MITE, it is very relevant that small subsystems have thermal (highly mixed) density matrices, but the whole system may well be in a pure state.)

Finally, the concepts of MATE and MITE show that thermal equilibrium \emph{can} be defined in a way that allows a system in a pure state to be in thermal equilibrium. At the same time, they also allow a system in a mixed state $\hat\rho$ to be in thermal equilibrium, without requiring that $\hat\rho$ be close to $\hat\rho^{(\beta)}$ or $\hat\rho^{\mc}$.
For example, even if the system is entangled with another system, and its state $\hat\rho$ is not pure, it could be much less mixed than $\hat\rho^{\mc}$; e.g., $\hat\rho$ could have rank 2 (i.e., could be a mixture of 2 pure states). 

Another subtlety in the quantum case arises from superpositions of macroscopically different states, such as Schr\"odinger's cat states. Here, our investigation touches upon the foundations of quantum mechanics. For the purposes of this paper, however, we can leave this problem aside.

\section{Thermal Equilibrium in Classical Mechanics}
\label{sec:classical}

A definition of thermal equilibrium for a classical system in a pure state amounts to the specification of a set of phase points that we regard as being in thermal equilibrium; that is, a subset $\Gamma_{\eq}$ of phase space $\Gamma$. Such a set $\Gamma_{\eq}$ has been defined by Boltzmann \cite{B96,Gol99} as follows. Consider a collection of macro variables $M_j$, $j=1,\ldots, K$; each of them can be regarded as a function on phase space, $M_j:\Gamma\to\RRR$. Since macro measurements have limited accuracy (say, $\Delta M_j>0$), we want to think of the $M_j$ as suitably coarse-grained with a discrete set of values, say, $\{k\Delta M_j: k\in \ZZZ\}$. Then two phase points $X_1,X_2\in\Gamma$ will look macroscopically the same if $M_j(X_1)=M_j(X_2)$ for all $j=1,\ldots, K$. In this way, the collection of functions $\{M_1,\ldots,M_K\}$ defines a partition of phase space $\Gamma$ into equivalence classes
\be\label{GammaMdef}
\Gamma_\nu = \Bigl\{X\in \Gamma: M_j(X)=\nu_j \:\forall j\Bigr\}\,,
\ee
one for every macro-state $\nu=(\nu_1,\ldots,\nu_K)$ described by the list of values of all $M_j$; we call $\Gamma_\nu$ a \emph{macro-state}. Some of the $\Gamma_\nu$ represent thermal equilibrium. 

\begin{figure}[h]
\begin{center}
\begin{minipage}{50mm}
\includegraphics[width=60mm]{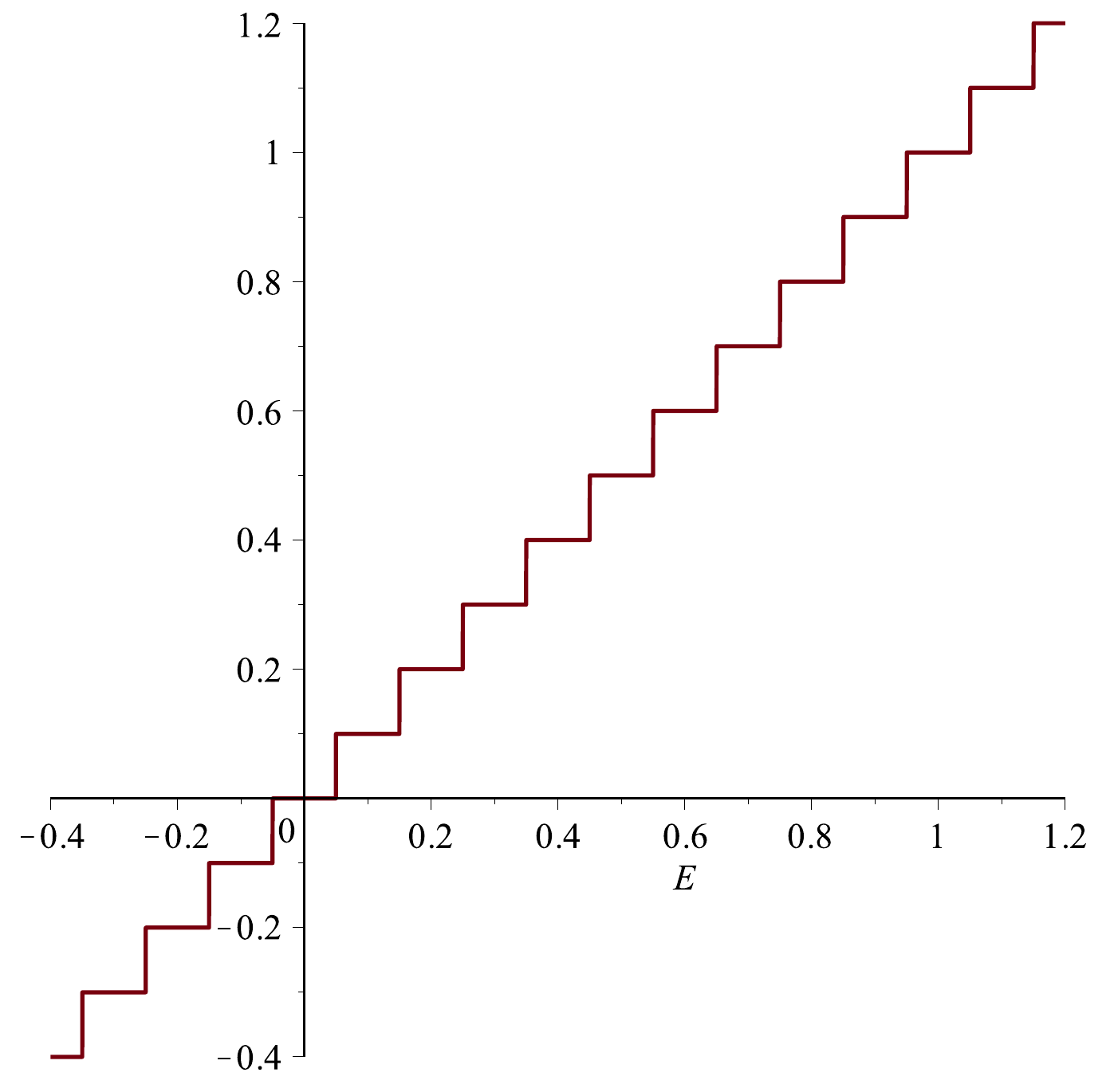}
\end{minipage}
\end{center}
\caption{Coarse graining function $f$ with $\Delta E = 0.1$}
\label{fig:coarse}
\end{figure}

\begin{figure}[h]
\begin{center}
\begin{minipage}{50mm}
\includegraphics[width=60mm]{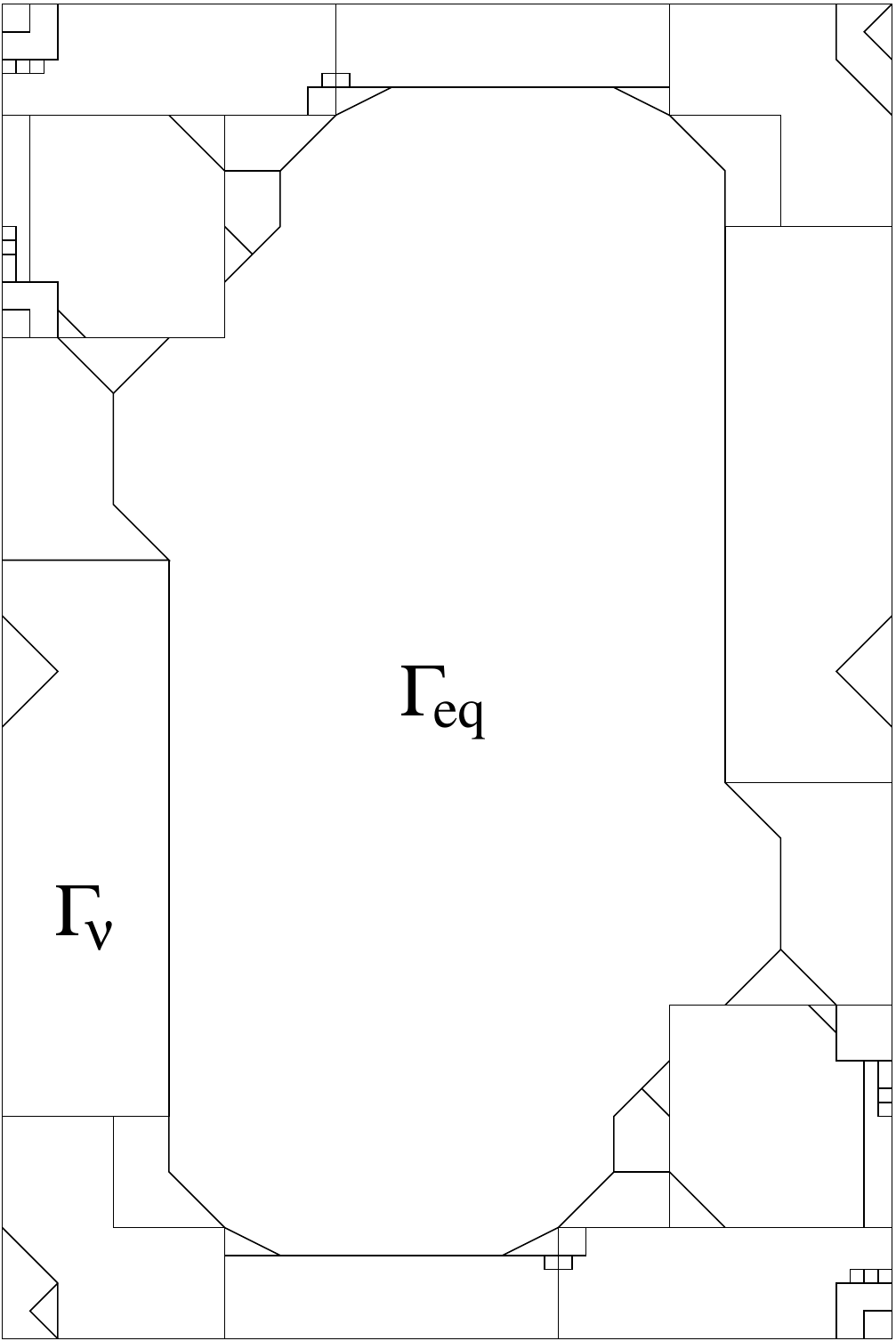}
\end{minipage}
\end{center}
\caption{Schematic representation of the partition of an energy shell $\Gamma_{\mc}$ in the phase space of a macroscopic classical system into subsets $\Gamma_\nu$ corresponding to different macro-states $\nu$. One of the subsets, $\Gamma_{\eq}$, contains more than 99.99\%\ of the volume (not drawn to scale) and corresponds to thermal equilibrium.}
\label{fig:phasespace}
\end{figure}

More specifically, since a coarse-grained version of the energy is usually among the macro variables, say $M_1(X) =  f\bigl(H(X)\bigr)$ with coarse-graining function $f(E) = [E/\Delta E]\, \Delta E$ and $[x]$ denoting the nearest integer to $x\in\RRR$ (see Figure~\ref{fig:coarse}), every macro-state $\Gamma_\nu$ belongs to a particular micro-canonical energy shell $\Gamma_{\mc}$,
so that $\Gamma_{\mc}$ is partitioned into macro-states $\Gamma_\nu$ (see Figure~\ref{fig:phasespace}). In most macroscopic systems (see Section~\ref{sec:exceptional} for a discussion of exceptions), there is, for every energy shell $\Gamma_{\mc}$, one macro-state $\Gamma_\nu=\Gamma_{\eq}$ that contains most of the phase space volume of $\Gamma_{\mc}$; see, e.g., \cite{Lan,Gol99,GL,L07}. A realistic value of the size of $\Gamma_{\eq}$ is
\be\label{Gammaeqsize}
\frac{\vol\Gamma_{\eq}}{\vol\Gamma_{\mc}} \approx 1-\exp(-10^{-15}N)\,,
\ee
where $\vol$ denotes the $6N$-dimensional phase space volume and $N$ is the number of degrees of freedom (or of particles) of the system; this estimate is derived in Section~\ref{sec:MATE} (having in mind a system that is macroscopically large).

Since the phase point $X(t)$ cannot leave the energy shell, and since phase space volume is conserved by Liouville's theorem, most $X\in\Gamma_{\eq}$ stay during their time evolution in $\Gamma_{\eq}$ for a long time (in fact, usually for an \emph{extraordinarily} long time), though not forever. Then, this set $\Gamma_{\eq}$ is the \emph{thermal equilibrium subset} for energy $E$, and the system \emph{is in thermal equilibrium} whenever $X(t)\in \Gamma_{\eq}$.

There is some arbitrariness in the choice of the functions $M_j$. 
As a consequence, there is also some arbitrariness about which set exactly $\Gamma_{\eq}$ is. The attitude of Boltzmann's followers (including the authors) is that this arbitrariness is unproblematical, as any reasonable choice of $\Gamma_{\eq}$ will take up most of the volume of $\Gamma_{\mc}$. Rather, this arbitrariness makes it evident that there is no reason to expect a unique criterion for exactly which phase points are in thermal equilibrium, just as there is no unique criterion for exactly which strings of 0's and 1's should count as ``purely random-looking.''

\section{Thermal Equilibrium in Quantum Mechanics}
\label{sec:quantum}

As already mentioned, unlike in classical mechanics, in quantum mechanics we need to consider two different notions of thermal equilibrium, which we describe in turn in the following two subsections.

\subsection{Macroscopic Thermal Equilibrium}
\label{sec:MATE1}

For quantum mechanics, a construction analogous to the subdivision of $\Gamma_{\mc}$ into $\Gamma_\nu$'s (Figure~\ref{fig:phasespace}) goes back to von Neumann \cite{vN29,GLTZ10} and, in a preliminary form, to Einstein \cite{Ein14}. Let
\be
\SSS(\Hilbert)=\bigl\{\psi\in\Hilbert: \|\psi\|=1\bigr\}
\ee
denote the unit sphere in Hilbert space. Consider a collection of macro observables, corresponding to self-adjoint operators $\hat{M}_j$, $j=1,\ldots,K$, on $\Hilbert$. These can be based on a
partition of the system's available volume $\Lambda\subset \RRR^3$ into cells $\Lambda_i$ that are small on the macro scale but
still large enough to each contain a large number of degrees of freedom. Examples of natural choices of $\hat M$'s are, for each cell, the number of
particles of each type, the energy of the cell, its momentum, and/or its magnetization. 
Again, we think of each $\hat{M}_j$ as suitably coarse-grained, so that its eigenvalues are separated by gaps whose magnitude corresponds to the inaccuracy of macro measurements. For example, the Hamiltonian $\hat{H}$ of a macroscopic system usually has eigenvalues separated by gaps much, much smaller than the macro energy inaccuracy $\Delta E$, so coarse graining at coarseness $\Delta E$, as in $\hat{M}_1 = f(\hat{H})$ with $f(E) = [E/\Delta E]\, \Delta E$ as before (see Figure~\ref{fig:coarse}), leads to a high degree of degeneracy of each eigenvalue. 

As von Neumann \cite{vN29} argued, the $\hat{M}_j$ can be taken to commute with each other,\footnote{For a different but closely related notion of thermal equilibrium,
proposed by Tasaki, see Section~\ref{sec:TMATE}. In this approach one avoids the necessity of rounding the macro
observables to make them commute. This approach adds support to there always being a dominant
macrostate. It is however not so convenient for discussing the joint values of the macro observables,
especially the nonequilibrium values.
} by changing them if necessary, in addition to the coarse-graining, in a way that is negligible on macro scales \cite{GLMTZ09b,GLTZ10,Y}. Then the simultaneous diagonalization of the $\hat{M}_j$ provides a decomposition of Hilbert space into a sum of orthogonal subspaces $\Hilbert_\nu$,
\be\label{decomp}
\Hilbert= \bigoplus_{\nu} \Hilbert_\nu\,,
\ee
where $\nu=(\nu_1,\ldots,\nu_K)$, and $\Hilbert_\nu$ is the joint eigenspace of the $\hat{M}_j$ with eigenvalues $\nu_j$. We call the $\Hilbert_\nu$ \emph{macro-spaces}, as they are the analogs of the $\Gamma_\nu$ and correspond to different macro-states. If $\hat{M}_1$ is again the coarse-grained energy, then its eigenspaces are the micro-canonical energy shells $\Hilbert_{\mc}$, which are also decomposed by a subcollection of $\Hilbert_\nu$'s. 
In general, one macro-space in each $\Hilbert_{\mc}$ has most of the dimension of $\Hilbert_{\mc}$, and this is the \emph{thermal equilibrium subspace} $\Hilbert_{\eq}$. In analogy to the classical case, a realistic value of the ratio of dimensions is (see Section~\ref{sec:MATE})
\be\label{Hilberteqsize}
\frac{\dim\Hilbert_{\eq}}{\dim \Hilbert_{\mc}} \approx 1-\exp(-10^{-15}N)\,.
\ee

We choose a suitably small tolerance $\delta>0$ and say that a system with state $\hat\rho$ is in \emph{macroscopic thermal equilibrium (MATE)} if and only if 
\be\label{MATEdef}
\tr(\hat\rho \hat P_{\eq}) > 1-\delta\,.
\ee
We also write MATE for the set of all $\hat\rho$'s in $\Hilbert_{\mc}$ satisfying this condition, as well as for the set of all pure states $\psi\in\SSS(\Hilbert_{\mc})$ such that $\hat\rho=\pr{\psi}$ satisfies \eqref{MATEdef}. 
A definition of thermal equilibrium along these lines was used, e.g., in \cite{Gri,Pen04,RDO08,GLMTZ09b,Tas10,GHT13b,GHT14a,GHT14b}. 
We look at realistic values of $\delta$ in Section~\ref{sec:MATE}.

\bigskip

\noindent{\bf Remarks.}
\begin{enumerate}
\setcounter{enumi}{\theremarks}

\item\label{rem:mostMATE} \emph{Most pure states in the energy shell are in MATE.} We note that this statement is also true of MBL systems. As a precise version of the statement, suppose 
that one of the macro-spaces, $\Hilbert_{\eq}$, is dominant,
\be\label{dominant}
\frac{\dim \Hilbert_{\eq}}{\dim \Hilbert_{\mc}} > 1-\varepsilon
\ee
with $0< \varepsilon \ll \delta$. Then
\be\label{sizeMATE}
u_{\mc}(\MATE)> 1-\frac{\varepsilon}{\delta} \approx 1\,,
\ee
where $u_{\mc}$ is the normalized uniform (surface area) measure on $\SSS(\Hilbert_{\mc})$. 

Indeed, 
\begin{align}
\int\limits_{\SSS(\Hilbert_{\mc})} \!\!\! u_{\mc}(d\psi)\, \scp{\psi}{\hat P_{\eq}|\psi}
&=\tr (\hat\rho^{\mc} \hat P_{\eq})\label{avgpsieqpsi}\\
&=\frac{\dim \Hilbert_{\eq}}{\dim \Hilbert_{\mc}} > 1-\varepsilon\,,
\end{align}
but the average of $f(\psi)=\scp{\psi}{\hat P_{\eq}|\psi}$ could not be that high if no more than $1-\varepsilon/\delta$ of all $\psi$'s had $f(\psi)>1-\delta$.

In practice, as an order of magnitude, 
\be\label{eta}
\varepsilon< 10^{-10^5}\,,
\ee
(as follows from \eqref{Hilberteqsize} for $N>3\times 10^{20}$) and $\delta$ can be taken to be $\sqrt{\varepsilon}$, which is still comparable to $10^{-10^5}$; then, according to \eqref{sizeMATE}, the fraction of states outside of MATE is also $\leq\sqrt{\varepsilon}$.

\item\label{rem:mosteigenMATE} 
\emph{Most eigenstates of $\hat{H}$ are in MATE.} In fact, for any orthonormal basis $\{b_1,\ldots,b_D\}$ of $\Hilbert_{\mc}$, at least the fraction $1-\varepsilon/\delta$ (close to 1 since $\varepsilon \ll \delta$) of all basis vectors are in $\MATE$, since 
\be
\frac{1}{D}\sum_{i=1}^D \scp{b_i}{\hat P_{\eq}|b_i} 
= \frac{1}{D} \tr(\hat P_{\eq}) > 1-\varepsilon\,.
\ee
Thus, for example, also for Hamiltonians exhibiting many-body localization, most eigenstates are in MATE. 

\item \emph{Most mixed states in the energy shell are in MATE.} In fact, this is the case relative to \emph{any} unitarily invariant distribution, uniform or not, over the density matrices in $\Hilbert_{\mc}$. In other words, suppose that $\hat\rho=\sum_\alpha p_\alpha |b_\alpha\rangle \langle b_\alpha|$ is chosen randomly, with the eigenbasis $\{b_1,\ldots,b_D\}$ uniformly distributed over all orthonormal bases of $\Hilbert_{\mc}$ (corresponding to the Haar measure over the unitary group) and the eigenvalues $p_1,\ldots,p_D$ independent of $b_1,\ldots,b_D$ with any joint distribution on the set defined by the conditions $0\leq p_\alpha\leq 1$ and $\sum_\alpha p_\alpha =1$; then $\hat\rho\in \MATE$ with probability near 1. 

Indeed,
\be
\tr(\hat\rho \hat P_{\eq}) = \sum_{i=1}^D p_i \scp{b_i}{\hat P_{\eq}|b_i}\,,
\ee
which always lies between 0 and 1. If we average this quantity over the eigenbasis, we obtain $\sum_i p_i (\dim \Hilbert_{\eq}/\dim \Hilbert_{\eq})> 1-\varepsilon$ by \eqref{avgpsieqpsi}. Since a quantity between 0 and 1 whose average is close to 1 must be close to 1 with high probability, we find that, in fact, even in the subset of $\hat\rho$'s with fixed eigenvalues, most $\hat\rho$'s are in MATE, and a fortiori so if the eigenvalues are randomized.
\end{enumerate}
\setcounter{remarks}{\theenumi}

\subsection{Microscopic Thermal Equilibrium}
\label{sec:MITE1}

The concept of microscopic thermal equilibrium (MITE) is inspired by \emph{canonical typicality}, the observation \cite{GMM04,PSW05,PSW06,GLTZ06,Sug07} that for any not-too-large subsystem $S$ and most wave functions $\psi$ in the energy shell $\Hilbert_{\mc}$,
\be\label{rhopsirhomc}
\hat\rho^\psi_S \approx \hat\rho^{\mc}_S\,,
\ee
where
\be\label{rhoSdef}
\hat\rho^\psi_S= \tr_{S^c} \pr{\psi}
\ee
is the reduced density matrix of $S$ obtained by tracing out the complement $S^c$ of $S$, and
\be
\hat\rho^{\mc}_S= \tr_{S^c} \hat\rho^{\mc}
\ee
with $\hat\rho^{\mc}$ the micro-canonical density matrix as in \eqref{hatrhomcdef}.
If $S$ is small enough then
\be\label{mcS=betaS}
\hat\rho^{\mc}_S \approx \hat\rho^{(\beta)}_S
\ee
for suitable $\beta>0$, where the right-hand side is the partial trace,
$\hat\rho_S^{(\beta)}= \tr_{S^c} \, \hat\rho^{(\beta)}$,
of the canonical density matrix
$\hat\rho^{(\beta)}$ of the whole system as in \eqref{hatrhobetadef}.
Let $\ell_0$ be the largest length small enough so that \eqref{mcS=betaS} holds for every subsystem $S$ with diameter $\leq \ell_0$. As a consequence of \eqref{mcS=betaS}, for small $S$, $\hat\rho^\psi_S \approx \hat\rho^{(\beta)}_S$. Hence, it does not matter whether one starts from $\hat\rho^{\mc}$ or $\hat\rho^{(\beta)}$ (this fact is a version of equivalence of ensembles), and we will call either one the canonical or thermal density matrix for $S$.\footnote{The density matrix $Z^{-1}_S\exp(-\beta \hat{H}_S)$ with $\hat H_S$ the Hamiltonian of $S$
is sometimes called the canonical or thermal density matrix for $S$; it agrees with $\hat\rho_S^{(\beta)}$ if the interaction between $S$ and its complement can be neglected. If the interaction cannot be neglected, then $\hat\rho_S^{(\beta)}$ is the correct density matrix to use.} 
As a consequence, also a micro observable $\hat{A}$ concerning a small subsystem $S$ behaves ``thermally'' in the sense that if we were to make a quantum measurement of $\hat{A}$ then the probability distribution over its eigenvalues would agree with the thermal distribution, defined by $\hat\rho^{\mc}_S$ (or, equivalently, by $\hat\rho^{\mc}$ or $\hat\rho^{(\beta)}$). 

For a system in a mixed state $\hat\rho$, we write $\hat\rho_S=\tr_{S^c} \hat\rho$ for the reduced state of $S$. If $\hat\rho$ is such that
\be\label{MITEdef}
\hat\rho_S\approx \hat\rho^{\mc}_S
\ee
for every subsystem $S$ corresponding to a spatial region of diameter $\leq \ell_0$ (as defined after \eqref{mcS=betaS}), we say that the system is in \emph{microscopic thermal equilibrium (MITE)}. We also use the name $\MITE$ for the set of $\hat\rho$'s in $\Hilbert_{\mc}$ that fulfill this condition, as well as for the set of pure states $\psi\in \SSS(\Hilbert_{\mc})$ that fulfill this condition. 
A concept along these lines was used, e.g., in \cite{Rei08,LPSW08,RS12,Lych,NH15}.

As a precise version of \eqref{MITEdef}, we may take the condition
\be
\|\hat\rho_S- \hat\rho^{\mc}_S\|< \varepsilon\,,
\ee
where $\varepsilon\ll 1$ is a chosen tolerance and $\|\cdot\|$ means the trace norm, defined by
\begin{equation}\label{tracenormdef}
  \|M\| = \tr |M| = \tr \sqrt{M^* M}\,.
\end{equation}

\bigskip

\noindent{\bf Remarks.}
\begin{enumerate}
\setcounter{enumi}{\theremarks}

\item 
\emph{In classical mechanics there is no analog of MITE for pure states.} Indeed, a classical system in a pure state is represented by a point $X$ in phase space, that is, by a list of the positions and momenta of all particles. For a subsystem $S$, be it defined as consisting of the particles numbered 1 through 100 or as the particles in a certain region $\R$ of the available volume in $\RRR^3$, its state is then given by the list of positions and momenta of the particles in $S$, i.e., by a point $X_S$ in the phase space of $S$ that is determined by $X$. Thus, the state of $S$ is itself pure and never close to $\rho^{(\beta)}$. While a notion of MITE is not available for pure states in classical mechanics, a notion of MATE is, as described in Section~\ref{sec:classical} above.

\item \label{rem:subsub} \emph{Subsubsystem property.}
If $\hat\rho_S \approx \hat\rho^{\mc}_S$ for some subsystem $S$ then the same is true for every smaller subsystem $S'$ contained in $S$, 
just by taking another partial trace on both sides of the approximate equation $\hat\rho_S \approx \hat\rho^{\mc}_S$. 
As a consequence, for a system to be in MITE it suffices that $\hat\rho_S\approx \hat\rho^{\mc}_S$ for a few subsystems $S=S_i$ corresponding to spatial regions (possibly of diameter $>\ell_0$) such that every region of diameter $\ell_0$ is contained in one of these regions. 

\item\label{rem:can} \emph{Most pure states in the energy shell are in MITE} (even for MBL systems). The basis of this fact is canonical typicality \cite{GMM04,PSW05,PSW06,GLTZ06,Sug07}, which can be understood as an instance of the following mathematical proposition \cite{Sug07}: Let $\Hilbert_R$ be any subspace of $\Hilbert$ of dimension $d_R$ (we will later set $\Hilbert_R=\Hilbert_{\mc}$), let $\hat P_R$ be the projection to $\Hilbert_R$ and $\hat\rho_R=\hat P_R/d_R$; let $\Psi$ be drawn randomly according to $u_R$, the uniform distribution over $\SSS(\Hilbert_R)$. Then, for any operator $\hat A:\Hilbert\to\Hilbert$,
\be
 \EEE \scp{\Psi}{\hat A|\Psi}= \tr(\hat A\hat\rho_R)
\ee 
and 
\be
\Var \scp{\Psi}{\hat A|\Psi} \leq \frac{V_{\hat A}(\hat\rho_R)}{d_R+1}\,,
\ee 
where 
\begin{align}
V_{\hat A} (\hat\rho)
&:= \tr\Bigl[ \Bigl(\hat A-\tr(\hat A\hat \rho) \Bigr)^* \Bigl(\hat A-\tr(\hat A\hat \rho) \Bigr) \hat\rho \Bigr]\\
&= \tr(\hat A^*\hat A\hat \rho)-|\tr(\hat A\hat \rho)|^2\,.
\end{align}

This proposition follows by a little calculation from the fact \cite{vN29,Ull64,GMM04,GLMTZ15} that the coefficients $c_\alpha$ relative to any orthonormal basis $\{\phi_\alpha\}$ of a random vector $\Psi=\sum_\alpha c_\alpha \phi_\alpha$ that is uniformly distributed over the unit sphere in some Hilbert space $\Hilbert$ of dimension $d$ have the following moments: The first and third moments vanish, the second moments are
\be
\EEE(c_\alpha^* c_\beta) = \frac{\delta_{\alpha\beta}}{d}\,,
\ee
and the only non-vanishing fourth moments are
\be
\EEE\Bigl( |c_\alpha|^2 |c_\beta|^2\Bigr) = \frac{1+\delta_{\alpha\beta}}{d(d+1)}\,.
\ee

The above proposition yields together with the Chebyshev inequality that for any operator $\hat A$ and any $\varepsilon>0$, 
\be
u_R \Bigl\{\psi\in\SSS(\Hilbert_R): \bigl|\scp{\psi}{\hat A|\psi}-\tr(\hat A\hat\rho_R)\bigr| \leq \varepsilon \Bigr\} \geq 1-\frac{V_{\hat A}(\hat\rho_R)}{\varepsilon^2(d_R+1)}\,.
\ee
Now suppose that $\Hilbert=\Hilbert_1 \otimes \Hilbert_2$ with $\dim \Hilbert_1=d_1$. By considering $\hat A$'s that act only on $\Hilbert_1$, one can further conclude through a little computation that 
\be
u_R \Bigl\{\psi\in\SSS(\Hilbert_R): \bigl\| \hat\rho_1^\psi-\tr_2 \hat\rho_R\bigr\| \leq \varepsilon \Bigr\} \geq 1-\frac{d_1^4}{\varepsilon^2 d_R}\,.
\ee
That is, when 
\be\label{mostMITEcond2}
d_1 \ll d_R^{1/4}
\ee
then $\hat\rho_1^\psi\approx \tr_2 \hat\rho_R$ for most $\psi\in\SSS(\Hilbert_R)$. [In fact, the tighter error estimate of Popescu et al.~\cite{PSW05,PSW06} (see Section~\ref{sec:MITE} below) yields that $d_R^{1/4}$ in \eqref{mostMITEcond2} can be replaced by $d_R^{1/2}$ (but not any larger exponent).] Now for $\Hilbert_R=\Hilbert_{\mc}$ and $\Hilbert_1$ the Hilbert space of a subsystem $S$, this amounts to canonical typicality.

What if $d_1=\infty$? If we are considering a system of $N_1+N_2$ spins (finitely many), then we do not encounter this problem, as the Hilbert spaces $\Hilbert_1$ and $\Hilbert_2$ have finite dimension $2^{N_1}$ and $2^{N_2}$. But if we are considering particles in a region $\Lambda=\Lambda_1\cup \Lambda_2\subset \RRR^3$, then both $\Hilbert_1$ and $\Hilbert_2$ have infinite dimension, although $\Hilbert_{\mc}$ has finite dimension, provided that $\Lambda$ has finite volume (as then there are only finitely many energy levels below $E$). That is why $d_1=\infty$ can occur. So, what if $d_1=\infty$? Effectively, only finitely many dimensions of $\Hilbert_1$ are relevant to $\Hilbert_{\mc}$: Let $\tilde\Hilbert_1$ be the span of the eigenvectors of $\tr_2 \rho_{\mc}$ with the largest $n$ eigenvalues; take $n$ large enough so that the sum of these eigenvalues is close to 1. Then $\tilde\Hilbert_1$ and an analogously constructed $\tilde\Hilbert_2$ can play the roles of $\Hilbert_1$ and $\Hilbert_2$ in the above reasoning.

Concerning the size of $S$, it follows from \eqref{mostMITEcond2} that canonical typicality still holds if the size of $S$ is almost one quarter of the size of the whole; in fact, by the tighter estimate $d_R^{1/2}$, almost one half (see Section~\ref{sec:MITE} for elaboration). If the diameter of the whole is greater than $4\ell_0$, then a moderate number (such as 8 for a cube) of nearly-half-size subsystems will contain any spatial region of diameter $\leq \ell_0$. By the subsubsystem property, we obtain that most $\psi\in\SSS(\Hilbert_{\mc})$ simultaneously satisfy $\hat\rho_S^\psi\approx \hat\rho_S^{\mc}$ for every region $S$ of diameter $\leq \ell_0$. That is, most $\psi\in\SSS(\Hilbert_{\mc})$ are in MITE.
\end{enumerate}
\setcounter{remarks}{\theenumi}

\subsection{Relation Between MATE and MITE}

\subsubsection{General Framework of MATE and MITE as Referring to Different Observables}
\label{sec:framework}

MITE and MATE are special cases of the following scheme: Given a set $\sA$ of observables, a state $\hat\rho$ in $\Hilbert_{\mc}$ is in \emph{thermal equilibrium relative to $\sA$} if and only if for every $\hat A\in\sA$, the probability distribution over the spectrum of $\hat A$ defined by $\hat\rho$ is approximately equal to that defined by $\hat\rho^{\mc}$. For $\sA=\sA_{\MATE}=\{\hat M_1,\ldots,\hat M_K\}$, one obtains MATE, and MITE is obtained for $\sA=\sA_{\MITE}=\cup_S\sA_S$ with the union taken over all spatial regions $S$ of diameter $\leq \ell_0$ and $\sA_S$ the set of all self-adjoint operators on $\Hilbert_S$, more precisely
\be\label{sASdef}
\sA_S =  \Bigl\{ \hat A_0\otimes \hat I_{S^c}: \hat A_0 \text{ self-adjoint on }\Hilbert_{S} \Bigr\}
\ee
(with $\hat I$ the identity operator and $S^c$ again the complement of $S$). Indeed, the condition $\hat\rho_S\approx \hat\rho^{\mc}_S$ is equivalent to $\tr(\hat\rho\hat P) \approx \tr(\hat\rho^{\mc} \, \hat P)$ for every projection of the form $\hat P = \hat P_0 \otimes \hat I_{S^c}$ with $\hat P_0$ a projection in $\Hilbert_S$.

In this sense, MATE means thermal equilibrium relative to the macro observables, whereas MITE is thermal equilibrium relative to all observables concerning any $S$ of diameter $\leq \ell_0$. The latter observables include those of a more microscopic and local nature. 

Yet another choice of $\sA$ has been considered by Reimann \cite{Rei15b}, who took $\sA$ to contain one or a few \emph{typical} observables (instead of macroscopic or local ones).

\subsubsection{MITE Implies MATE for Macroscopic Systems}
\label{sec:MITEimpliesMATE}

Since most $\psi\in\SSS(\Hilbert_{\mc})$ are in both MATE and MITE (see Remarks~\ref{rem:mostMATE} and \ref{rem:can} above), it follows further that most states in $\MITE$ lie also in $\MATE$ and vice versa. (Indeed, if $99\%$ of all states lie in MITE, and $99\%$ of all states lie in MATE, then at least the fraction $1-1/99$ of all states in MITE lie in MATE, and at least the fraction $1-1/99$ of all states in MATE lie in MITE.) 
In fact, more is true: \emph{All} states in MITE lie also in MATE \cite{GHLT15a}. 

Indeed, since macro observables are sums or averages of local observables over spatial cells (say, of length $L$), it follows from Section~\ref{sec:framework}, as soon as $L\leq \ell_0$, that states $\psi$ that display thermal behavior for micro observables (i.e., lead to the same probability distribution over the spectrum of the observable as $\hat\rho^{\mc}$) also display thermal behavior for macro observables. And these $\psi$ include those in MITE. The condition $L\leq \ell_0$ means that $\hat\rho^\psi_S \approx \hat\rho^{\mc}_S$ at least up to the length scale of the macro observables, which is commonly the case; e.g., for a cubic meter of gas at room conditions, we can realistically take $L \approx 10^{-4}$ m and $\ell_0 \approx 10^{-3}$ m.

\bigskip

\noindent {\bf Example 1.} \emph{A simple example of a state in MATE that is not in MITE.} Consider a system of $N\gg 1$ non-interacting spins-1/2, $\Hilbert=(\CCC^2)^{\otimes N}$, with $\hat H=0$ so that $\Hilbert_{\mc}=\Hilbert$ and $\hat\rho^{\mc}=2^{-N}\hat I$, in a pure product state $\psi=\otimes_i \psi_i$. Divide the $N$ spins into $m$ groups (``cells'') $\Lambda_j$ of $n\gg 1$ spins, so that $nm=N$, and take $\hat M_j$ to be a coarse-grained version of $\sum_{i\in\Lambda_j} \hat\sigma_i^z$, the total magnetization of $\Lambda_j$ in the $z$-direction. Then the thermal equilibrium value of $\hat M_j$ is $\tr(\hat\rho^{\mc}\hat M_j)=0$, so $\Hilbert_{\eq}=\bigcap_j \,\mathrm{kernel}(\hat M_j)$ (where kernel means the eigenspace with eigenvalue 0), and a typical pure product state $\psi$ lies in MATE. To see that $\psi$ does not lie in MITE, note that for a single spin at site $i$, $S=\{i\}$, 
\be\label{H=0}
\hat\rho^{\mc}_S=\tfrac 12 \hat I_i\text{ whereas }\hat\rho_S^\psi= |\psi_i\rangle\langle\psi_i|\,,
\ee
so the two density matrices are not close to each other.

\section{Dynamical Approach to Thermal Equilibrium}
\label{sec:approach}

We say that $\hat\rho$ \emph{approaches} MITE/MATE if $\hat\rho_t= e^{-i\hat{H}t} \,\hat\rho\, e^{i\hat{H}t}$ sooner or later reaches MITE/MATE and spends there most of the time in the long run. In many systems, all states in the energy shell approach thermal equilibrium in this sense, but there 
are some exceptional macroscopic classical and quantum systems for which many states do \emph{not} come to thermal equilibrium in any sense as time goes on. This is obviously the case for Example 1 above, but in fact there are more physically relevant systems (exhibiting MBL) which also have this property, as we explain below. 

A condition relevant to whether approach to thermal equilibrium occurs is the
\emph{eigenstate thermalization hypothesis (ETH)} \cite{Sre94,RDO08,GLMTZ09b,RS12}. The ETH can be formulated as the condition on $\hat H$ that all eigenstates of $\hat{H}$ are in thermal equilibrium. Corresponding to two kinds of thermal equilibrium, MITE and MATE, we have two versions of the ETH. Let us focus first on MATE-ETH.

\subsection{Approach to MATE}
\label{sec:approachMATE}

\emph{Under the MATE-ETH, all $\psi$ in the energy shell approach MATE.} Indeed \cite{GLMTZ09b}, writing $\overline{f(t)}=\lim_{T\to\infty} \frac{1}{T}\int_0^T f(t)\, dt$ for time averages, $\ket{\alpha}$ for the energy eigenstate with eigenvalue $E_\alpha$, and $\psi_t=e^{-i\hat Ht}\psi$,
\begin{align}
\overline{\scp{\psi_t}{\hat P_{\eq}|\psi_t}}
&= \sum_{\alpha,\beta} \scp{\psi}{\alpha} \: \overline{e^{iE_\alpha t} \scp{\alpha}{\hat P_{\eq}|\beta} e^{-iE_{\beta}t}} \: \scp{\beta}{\psi} \label{MATE-ETH-first}\\
&= \sum_{\alpha} \bigl| \scp{\psi}{\alpha} \bigr|^2  \scp{\alpha}{\hat P_{\eq}|\alpha}
\geq \sum_{\alpha} \bigl| \scp{\psi}{\alpha} \bigr|^2  (1-\delta)
=1-\delta\,,\label{MATE-ETH-last}
\end{align}
provided $\hat H$ is non-degenerate, i.e., $E_\alpha\neq E_{\beta}$ for $\alpha\neq \beta$ (using $\overline{e^{iEt}}=1$ if $E= 0$ and $=0$ otherwise).\footnote{In fact, the assumption of non-degeneracy can be dropped: If we number the eigenvalues as $E_\alpha$ with $E_\alpha\neq E_{\beta}$ for $\alpha\neq \beta$ and let $\ket{\alpha}$ denote the normalized projection of $\psi$ to the eigenspace of $E_\alpha$, then the calculation \eqref{MATE-ETH-first}--\eqref{MATE-ETH-last} still applies.} Since its time average is close to 1, $\scp{\psi_t}{\hat P_{\eq}|\psi_t}$ must be close to 1 for most $t$ in the long run.

Conversely, if the MATE-ETH is violated, then it is mathematically possible that no state outside MATE ever approaches MATE. For example, choose $\hat H$ so that every eigenstate is either in $\Hilbert_{\eq}$ or orthogonal to it. As we will see in Section~\ref{sec:mbl}, this happens in some MBL systems.

\subsection{Approach to MITE}
\label{sec:approachMITE}

The ideal gas provides an example of a system for which some states do not approach MITE. We now ask, Under which conditions will all or most $\psi$ approach MITE? There are several results \cite{RDO08,Rei08,LPSW08,Rei10}, all of which assume the MITE-ETH, and that the Hamiltonian is non-degenerate and has non-degenerate energy gaps, i.e.,
\be\label{noresonance}
E_{\alpha}-E_{\beta} \neq E_{\alpha'}-E_{\beta'} \text{ unless }
\begin{cases}\text{either } \alpha= \alpha' \text { and } \beta= \beta' \\
\text{or }\alpha=\beta \text{ and }\alpha'=\beta'\,,\end{cases}
\ee
a condition that is generically fulfilled. 

We note here two results, the first of which \cite{Rei08,LPSW08} asserts that if all energy eigenstates in $\Hilbert_{\mc}$ are in MITE, then most $\psi\in\SSS(\Hilbert_{\mc})$ will sooner or later reach MITE and spend most of the time in MITE in the long run. More precisely, those $\psi$ will behave this way for which the effective number of significantly participating energy eigenstates is much larger than $\dim \Hilbert_S$ for any small $S$. 

The second result \cite{RDO08} shows that \emph{all} (rather than most) $\psi$ will ultimately reach MITE and stay there most of the time, under two assumptions, first again that all energy eigenstates $|\alpha\rangle$ are in MITE, and second Srednicki's \cite{Sre96,Sre99} extension of the ETH to off-diagonal elements, i.e., that for $\hat A\in\sA_{\MITE}$ (as in Section~\ref{sec:framework}), 
\be
\scp{\alpha}{\hat A|\beta}\approx 0\text{ for }\alpha\neq \beta
\ee
(see also \cite{Rei15}). Indeed, if $\hat H$ is non-degenerate and all $\ket{\alpha}$ are in MITE, then
\begin{align}
\overline{\scp{\psi_t}{\hat A|\psi_t}}
&= \sum_{\alpha,\beta} \scp{\psi}{\alpha} \overline{e^{iE_\alpha t} \scp{\alpha}{\hat A |\beta} e^{-iE_\beta t}} \scp{\beta}{\psi}\\
&= \sum_\alpha \scp{\psi}{\alpha} \scp{\alpha}{\hat A|\alpha} \scp{\alpha}{\psi}\\
&\approx \tr(\hat\rho^{\mc} \hat A)\,.
\end{align}
Furthermore, a calculation using \eqref{noresonance} shows that
\be\label{timevariance}
\overline{\biggl( \scp{\psi_t}{\hat A|\psi_t} - \overline{\scp{\psi_t}{\hat A|\psi_t}} \biggr)^{\!2}}
=\sum_{\alpha\neq \beta} \bigl|\scp{\psi}{\alpha}\bigr|^2 \; \bigl|\scp{\alpha}{\hat A|\beta}\bigr|^2 \;
\bigl|\scp{\beta}{\psi}\bigr|^2\,,
\ee
and if $\bigl|\scp{\alpha}{\hat A|\beta}\bigr|<\varepsilon\ll 1$ for all $\alpha\neq \beta$, then the time variance \eqref{timevariance} is smaller than $\varepsilon^2$.  
It follows that, for most $t$ in the long run, $\scp{\psi_t}{\hat A|\psi_t} \approx \tr(\hat\rho^{\mc}\, \hat A)$ for any $\hat A\in\sA_{\MITE}$ (in particular projections), which yields that $\psi_t\in \MITE$ for most $t$ in the long run.

\section{Many-Body Localized Systems}
\label{sec:mbl}

There is no consensus on the definition of many-body localization \cite{GE15}.  For the purposes of this paper we will adopt the following definition:  A system with Hamiltonian $\hat H$ is many-body localized if all the eigenstates of $\hat H$ fail to be in MITE, this remains true under generic small local (in real space) changes to $\hat H$, and in each eigenstate almost all subsystems $S$ are ``localized'' with $\hat\rho_S$ having substantially lower entropy than at thermal equilibrium.

Many-body localized (MBL) systems are the one known example of many-body quantum systems that fail to thermalize under their own dynamics where this failure to thermalize remains under small generic local perturbations to the system's Hamiltonian. Since the approach to thermal equilibrium is connected to the properties of the energy eigenstates $\phi_\alpha$, it is of particular interest whether the $\phi_\alpha$ lie in MITE or MATE or neither.

\bigskip

\noindent{\bf Example 2.}
As a simple, and essentially trivial, example, consider a chain of $N$ non-interacting spins-1/2, each subject to a local random field:
\be\label{exMBL1}
\hat{H_2}=\sum_i h_i \hat\sigma_i^z ~,
\ee
where $i$ labels the spin, and $\hat\sigma_i^z$ is the Pauli operator for the $z$ component of spin $i$.  For specificity, let the local static random fields $h_i$ be independent and identically distributed, drawn from the uniform distribution on $-W<h_i<W$ with $W>0$. Let us consider an energy shell containing $E=0$, which in a sense corresponds to infinite temperature. 

The eigenstates of $\hat H_2$ are simply the simultaneous eigenstates of each $\hat\sigma_i^z$, all of which mutually commute and thus also commute with $\hat H_2$.  In this sense, we have a (trivially) integrable system, with a complete set of local conserved operators, the $\{\hat\sigma_i^z\}$.  We take as macro observables the $\hat M_j$ of Example 1 above, i.e., the coarse-grained total $z$-magnetization in each macro cell, along with the coarse-grained energy $\hat M_0=f(\hat H_2)$. 
So, $\Hilbert_{\eq}=\Hilbert_{\mc}\cap \bigcap_j \, \mathrm{kernel}(\hat M_j)$. Then, most of the eigenstates of $\hat H_2$ in $\Hilbert_{\mc}$ are in MATE (in fact, even in $\Hilbert_{\eq}$), with an energy near zero (for this energy shell) in all large subregions of the spin chain.  But there are also a few eigenstates where some large subregions have energies that deviate substantially from the thermal equilibrium value 0, and these are the eigenstates that are not in MATE. In fact, these eigenstates are orthogonal to $\Hilbert_{\eq}$ (i.e., as far from MATE as possible). So, the situation can be summarized by the statement that
\be\label{ineqperpeq}
\text{for every $\alpha$, either }\phi_\alpha\in \Hilbert_{\eq}\text{ or }\phi_\alpha \perp \Hilbert_{\eq}\,. 
\ee
Thus, MATE-ETH is violated as strongly as consistent with the mathematical fact (Remark~\ref{rem:mosteigenMATE} above) that always most eigenstates are in MATE. As a consequence of \eqref{ineqperpeq}, all states out of MATE stay out of MATE forever (no MATE-thermalization and, since MITE implies MATE, also no MITE-thermalization, the heat transport coefficients vanish), whereas all states in MATE stay in MATE forever (no fluctuations away from thermal equilibrium). 

It is moreover the case for $\hat H_2$ (but not relevant to thermalization) that every eigenstate $\phi_\alpha$ lies in some $\Hilbert_\nu$. In contrast, a typical Hamiltonian, say one whose eigenbasis in $\Hilbert_{\mc}$ was drawn uniformly among all orthonormal bases in $\Hilbert_{\mc}$, has eigenvectors $\phi_\alpha$ that are typical vectors relative to the uniform distribution over $\SSS(\Hilbert_{\mc})$, and by the phenomenon of \emph{normal typicality} \cite{vN29,GLMTZ10,GLTZ10}, $\|\hat P_\nu\phi_\alpha\|^2\approx \dim \Hilbert_\nu/\dim \Hilbert_{\mc}$, where $\hat P_\nu$ is the projection to $\Hilbert_{\nu}$; that is, $\phi_\alpha$ is spread out over \emph{all} $\Hilbert_{\nu}$; in fact, this is the case for \emph{all} $\alpha$ simultaneously \cite{vN29,GLMTZ10,GLTZ10}. Correspondingly, \cite{GLMTZ09b,GLMTZ10}, for a typical Hamiltonian every $\phi_\alpha$ has a component of size $1-\varepsilon = \dim\Hilbert_{\eq}/\dim\Hilbert_{\mc}$ in $\Hilbert_{\eq}$ and a component of size $\varepsilon$ orthogonal to $\Hilbert_{\eq}$; as a consequence, all $\phi_\alpha$ lie in MATE \cite{GLMTZ09b}. It is ironical that, although MATE is nearly the same as $\Hilbert_{\eq}$, almost none of the eigenstates can lie in $\Hilbert_{\eq}$ if all of them lie in MATE (and not all can lie in MATE if as many as possible of them lie in $\Hilbert_{\eq}$).

Several traits of the eigenstates of $\hat H_2$ are quite typical of MBL systems: The energy eigenstates $\phi_\alpha$ of MBL systems tend to have a short range of entanglement. That is, while they are not exactly product states, they are less entangled between neighboring lattice sites than random states $\psi$ (and thus less than for Hamiltonians with a random eigenbasis). They can be approximated as unentangled between different cells $\Lambda_j$. That is, the $\phi_\alpha$ of an MBL system can be approximated as a product of eigenstates of local energy, a situation of which Example 2 is a strict case. As a consequence, some eigenstates have a profile of cell energy that is very non-uniform, and they will not be in MATE, but will be approximately orthogonal to $\Hilbert_{\eq}$. 
In addition, for a generic interacting MBL system there are presumably also a small number of eigenstates that contain substantial components both in $\Hilbert_{\eq}$ and orthogonal to $\Hilbert_{\eq}$; this should happen when the profile of cell energy lies near the borderline of what should be considered uniform.

Let us have a look at MITE in Example 2. In our energy shell $E=0$, the thermal density matrix of a single spin at site $i$, $S=\{i\}$, is
\be
\hat\rho^{\mc}_S= \tfrac12 |\!\uparrow\rangle\langle \uparrow\!| 
+ \tfrac12 |\!\downarrow\rangle\langle \downarrow\!|\,,
\ee
analogously to the even more trivial Example 1 in Section~\ref{sec:MITE1}. However, also analogously to \eqref{H=0}, for every eigenstate $\phi$ of $\hat H_2$, due to the product structure of $\phi$, $\hat\rho_S^\phi=\hat\rho_i^\phi$ is either $|\!\!\uparrow\rangle\langle \uparrow\!\!|$ or $|\!\!\downarrow\rangle\langle \downarrow\!\!|$, so $\hat\rho_S^\phi$ is far from $\hat\rho^{\mc}_S$.
Thus, for this system, the MITE-ETH is false, in fact none of the eigenstates are in MITE.  [In the full spectrum, there are two exceptional eigenstates, namely the ground states of $\hat H_2$ and of $-\hat H_2$.  These two states are (trivially) in both MITE and MATE, as is always the case for non-degenerate ground states.  But we are not interested here in ground states.]

Also this situation is typical of MBL systems: It has been shown analytically \cite{Imbrie}, numerically \cite{PH10}, or perturbatively \cite{BAA06a,BAA06b,ros} for various MBL models that none, or almost none, of the eigenstates of $\hat H$ are in MITE---although most pure states, when consideration is not restricted to the energy eigenstates, are necessarily in MITE. {This is only to be expected considering that entanglement has short range in $\phi_\alpha$ of MBL systems, and entanglement is the mechanism behind MITE. So, a typical MBL system has most energy eigenstates in MATE but none in MITE; it is thus as far from the ETH as possible, in view of the mathematical fact (Remark~\ref{rem:mosteigenMATE}) that most energy eigenstates have to be in MATE.

Correspondingly to the failure of MITE-ETH, typical $\psi$'s in Example 2 do not approach MITE. For example (though not a typical one), if $\psi$ is initially a product state then it will forever remain one due to the form of \eqref{exMBL1}, and product states lack the entanglement needed for MITE, so $\psi$ never reaches MITE.

Now consider a pure state $\psi\in\Hilbert_{\mc}$ built out of the eigenstates $\phi_\alpha$ in MATE. Since it lies in a subspace in which the MATE-ETH is true, $\psi$ approaches MATE (see Section~\ref{sec:approachMATE}). Can $\psi$ be at all out of MATE (so that it is a non-thermal state that thermalizes)? For an ETH system, it is clear that the answer is yes, i.e., a non-MATE $\psi$ can be superposed out of MATE eigenstates, as all eigenstates $\phi_\alpha$ from $\Hilbert_{\mc}$ lie in MATE, and surely some $\psi\in\Hilbert_{\mc}$ are not in MATE, so they must be built of $\phi_\alpha$'s in MATE. It is equally clear that for Example 2 the answer is no, as the $\phi_\alpha$ in MATE lie in $\Hilbert_{\eq}$, so any superposition also lies in $\Hilbert_{\eq}$ and thus in MATE. So what about other MBL systems? This will be addressed by the next example.

\bigskip

\noindent{\bf Example 3.} 
Take the same Hilbert space and Hamiltonian as in Example 2, but now take the macro observables $\hat M_j$ to refer only to $x$-spin and not to $z$-spin, which leads to a different choice of $\Hilbert_{\eq}$. (This example is less serious because serious examples should have cell energies among their macro variables, and this example does not; but we consider it anyway.)

It is useful to consider the basis $\{b_\alpha\}$ of $\Hilbert$ consisting of products of $|\!\!\rightarrow\rangle$'s and $|\!\!\leftarrow\rangle$'s; those $b_\alpha$ that have approximately equally many $|\!\!\rightarrow\rangle$'s as $|\!\!\leftarrow\rangle$'s in every cell lie in $\Hilbert_{\eq}$, and the others are orthogonal to $\Hilbert_{\eq}$. It follows that every energy eigenstate $\phi_\alpha$ lies in MATE. As a consequence, every $\psi\in\Hilbert$ approaches MATE for this choice of macro variables. 

For example, $\psi=|\!\!\rightarrow\rangle^{\otimes N}$ is orthogonal to $\Hilbert_{\eq}$, and in particular not in MATE (and hence not in MITE). Since all spins precess at different frequencies due to the local random fields $h_i$ in \eqref{exMBL1}, the macroscopic $x$-magnetization relaxes to zero, and this $\psi$ approaches MATE, as it should. 

In view of this example of a dynamical relaxation of the $x$-magnetization, we may ask whether there are also pure states of macroscopically non-uniform cell energy distribution (i.e., non-uniform temperature profile, such as a temperature gradient) that relax to a uniform cell energy distribution. The answer is no, as it is clear from \eqref{exMBL1} that the spins do not interact and thus energy cannot be transported from one site to another. In fact, it follows that a $\psi$ with a non-uniform cell energy distribution must consist exclusively of energy eigenstates with the same cell energy distribution.

With respect to MITE, Example 3 behaves like Example 2 because MITE does not depend on the choice of the $\hat M_j$, which was the only difference between the two examples. That is, none of the $\phi_\alpha$ lie in MITE, and approach to MITE does not occur.

\bigskip

\noindent{\bf Example 4.} 
Take again the same Hilbert space and Hamiltonian as before, but now let us include among the $\hat M_j$ \emph{both} the $x$-spin and the $z$-spin on the macro level. That is, we include coarse-grained magnetization operators for each cell in the $x$- and the $z$-direction (a little adjusted so as to make them all commute). This is a natural choice that reflects better what can macroscopically be measured. 

Then, again, none of the energy eigenstates lie in MITE, and approach to MITE does not occur. Some $\phi_\alpha$ (those with a macroscopically non-uniform density of $|\!\!\uparrow\rangle$ factors) will be (approximately) orthogonal to $\Hilbert_{\eq}$ and thus clearly out of MATE; $|\!\!\rightarrow\rangle^{\otimes N}$ (or rather, its normalized projection to $\Hilbert_{\mc}$) will again be an example of a state out of MATE that approaches MATE. So, some non-equilibrium states thermalize, but a non-zero temperature gradient cannot relax. 

\bigskip

\noindent{\bf Example 5.} For a system that is less trivially localized, let us now add some nearest neighbor interactions to this spin chain model, as well as possibly a transverse field.  For example, Imbrie \cite{Imbrie} adds non-random Ising interactions and a transverse field:
\be
\hat{H_5}=\sum_i (J\hat\sigma_i^z\hat\sigma_{i+1}^z + \Gamma\hat\sigma_i^x + h_i \hat\sigma_i^z) ~.
\ee
For $W>0$ and $\Gamma$ small enough, he shows \cite{Imbrie} under plausible assumptions that this system remains fully many-body localized (although the precise definition of MBL that he uses differs from ours in ways that we expect are not important for the present discussion).  In this regime, for any small local perturbation of $\hat H_2$ one can define localized conserved operators $\hat\tau_i^z$ that all mutually commute and also commute with the resulting Hamiltonian $\hat H_5$ \cite{hno,ros}.  These operators $\hat\tau_i^z$ are made by ``dressing'' each $\hat\sigma_i^z$ with multi-spin operators that are localized near site $i$. This means that the norm of any such dressing typically falls off exponentially with the distance of the farthest spin used in the dressing and the probability of having strong long-range dressing also falls off exponentially with the distance.  In terms of these $\{\hat\tau_i^z\}$, the Hamiltonian of this more generic system can be written as \cite{hno}
\be
\hat{H_5}=\sum_i \tilde h_i \hat\tau_i^z + \sum_{i<j} J_{ij}\hat\tau_i^z\hat\tau_j^z + \sum_{i<j<k}K_{ijk}\hat\tau_i^z\hat\tau_j^z\hat\tau_k^z+ \ldots ~,
\ee
where $\tilde h_i$ is the local effective random field, and the interactions $J_{ij}$, $K_{ijk}$, etc. typically fall off exponentially with the distance between the two farthest operators involved, as does the probability of such a coupling being strong.

Although $\hat H_5$ has more interactions than $\hat H_2$, it is similarly integrable with a complete set of localized conserved operators, the $\{\hat\tau_i^z\}$.  And $\hat H_5$ has all the properties outlined above for $\hat H_2$, including some eigenstates that fail to be in MATE, and having all highly-excited eigenstates fail to be in MITE.  

It remains an open question whether or not all systems that are MBL have this structure, with a complete set of localized conserved operators.  No detailed description of how MBL would work otherwise has yet been proposed.

\section{Further Aspects of MITE and MATE}
\label{sec:overview}

\subsection{Quantitative MATE}
\label{sec:MATE}

In this section, we focus on the practical size of $\varepsilon$ and $\delta$ in \eqref{MATEdef} and \eqref{dominant}; that is, of $\varepsilon=1-\dim\Hilbert_{\eq}/\dim\Hilbert_{\mc}$ (or, classically, $\varepsilon=1-\vol\Gamma_{\eq}/\vol\Gamma_{\mc}$), and of the $\delta$ that quantifies how far $\tr(\hat\rho \hat P_{\eq})$ can deviate from 1 in MATE. 

First, we stated in \eqref{Gammaeqsize} and \eqref{Hilberteqsize} that
\be\label{MATEepsilon}
\varepsilon \approx \exp(-10^{-15}N)\,.
\ee
The estimate is similar in the classical and in the quantum case. To obtain it classically, we partition the available 3-volume $\Lambda\subset \RRR^3$ into $m$ (say, $10^9$) cells $\Lambda_i$ of equal volume, consider simply the configuration space $\Lambda^N$ instead of phase space, use the uniform distribution over $\Lambda^N$, and take the macro variables to be $M_j= [N_j/N\, \Delta M_j] \,\Delta M_j$, where $N_j$ is the number of particles in $\Lambda_j$ (and, say, $\Delta M_j=10^{-12}$). Then $M_j$ has equilibrium value $\nu_j^{\eq}=1/m$ (and the relative resolution is $\Delta M_j/\nu_j^{\eq}=10^{-3}$); the distribution of $N_j$ is binomial with parameters $N$ and $m^{-1}$ and thus, if $N$ is large, approximately Gaussian with parameters $\mu=N/m$ and $\sigma^2=m^{-1}(1-m^{-1})N\approx N/m$. For $M_j$ to deviate from its equilibrium value requires that $N_j$ deviates from $\mu$ by more than $N \, \Delta M_j$, i.e., by more than $\sqrt{mN}\, \Delta M_j$ standard deviations, which has probability less than $p:=\exp(-mN\, \Delta M_j^2)$. Since the $N_j$ are approximately independent, the probability that \emph{any} of the $M_j$ deviates from its equilibrium value is $mp$, which here is still of rough order of magnitude $\exp(-10^{-15}N)$. 

In this example we have chosen numbers appropriate for a truly macroscopic system (say, $N\geq 10^{20}$) and 
require  equilibrium values  to a rather high resolution in all of a rather large
number of cells. 
The numbers can reasonably be changed by many orders
of magnitude to consider much smaller systems, to demand equilibrium values to
different levels of precision and to divide the system into different numbers 
of cells.  At some point $N$ becomes too small to allow room for a reasonable definition of MATE.

We now turn to the question, How big should $\delta$ reasonably be chosen? Not too small, or else $\MATE$ will not contain the majority of $\SSS(\Hilbert_{\mc})$, and not too large, or else $\psi\in\MATE$ will have significant component orthogonal to $\Hilbert_{\eq}$ and will not mean much. That is,
\be
\varepsilon \ll \delta \ll 1\,.
\ee
Since a realistic value of $\varepsilon$ is $10^{-10^5}$ or smaller (taking $N\geq 10^{20}$), there is a lot of different possibilities for $\delta$. Since $\delta$ represents the maximal probability, in an ideal quantum measurement of $\hat P_{\eq}$ on $\psi\in\MATE$, of obtaining the outcome 0 and projecting $\psi$ to a subspace orthogonal to $\Hilbert_{\eq}$, we may want to choose this probability so small that we can expect never to observe such an outcome. Borel \cite[Chap.~6]{Borel} has argued that events with probability $<10^{-200}$ can be assumed to never occur in our universe, so we may want to choose $\delta<10^{-200}$. A natural choice is $\delta = \sqrt{\varepsilon\,}$.

\subsection{Quantitative MITE}
\label{sec:MITE}

As already mentioned, the statement that most $\psi\in\SSS(\Hilbert_{\mc})$ are in $\MITE$ is based on canonical typicality. A tighter estimate of canonical typicality than the one in Remark~\ref{rem:can} is provided by a 
theorem due to Popescu, Short, and Winter \cite{PSW05,PSW06}, which asserts that,
for any Hilbert spaces $\Hilbert_1$, $\Hilbert_2$ of dimensions $d_1$, $d_2$, any subspace $\Hilbert_R\subseteq \Hilbert_1 \otimes \Hilbert_2$ of dimension $d_R$, and any $\tilde\varepsilon>0$,
\begin{equation}\label{cantyp}
  u_R \left\{ \psi \in \SSS(\Hilbert_R): \Bigl\|\hat\rho_1^\psi -
  \tr_2 \hat\rho_R  \Bigr\| \geq \tilde\varepsilon + \frac{d_1}{\sqrt{d_R}}
  \right\} \leq 4 \exp\Bigl(-\frac{d_R\tilde\varepsilon^2}{18\pi^3}\Bigr)\,.
\end{equation}

Let us explain how this estimate can be applied. We can immediately consider several systems $S_1,\ldots,S_r$ simultaneously and ask, Under which conditions does the set
\be\label{MITElistdef}
M=\bigcap_{i=1}^r\Bigl\{ \psi\in\SSS(\Hilbert_{\mc}): \hat\rho_{S_i}^\psi \approx \hat\rho^{\mc}_{S_i} \Bigr\}
\ee
contain most wave functions? Here, we take
the relation $\hat\rho_{S_i}^\psi\approx \hat\rho_{S_i}^{\mc}$ to mean
\be\label{normepsilon}
\bigl\|\hat\rho_{S_i}^\psi-\hat\rho_{S_i}^{\mc} \bigr\|<\varepsilon
\ee
for some fixed $0<\varepsilon\ll 1$. (This $\varepsilon$ is independent of the quantity called $\varepsilon$ for MATE in \eqref{MATEepsilon} and \eqref{dominant}.)
Let $d_i=\dim\Hilbert_{S_i}$, $d_{\mc}=\dim\Hilbert_{\mc}$, and let $u_{\mc}$ denote the uniform probability distribution over $\SSS(\Hilbert_{\mc})$. From the theorem \eqref{cantyp} we obtain: If 
\be\label{mostMITEcondition}
d_i< \tfrac{1}{2} \varepsilon \sqrt{d_{\mc}} \text{ for all $i$}\,,
\ee
then
\be\label{mostMITE}
u_{\mc}(M) \geq 1- 4r \exp\Bigl(-\frac{d_{\mc}\varepsilon^2}{72\pi^3}\Bigr)\,.
\ee

Indeed, this follows by setting $\tilde\varepsilon=\varepsilon/2$, $\Hilbert_R=\Hilbert_{\mc}$, $\Hilbert_1=\Hilbert_{S_i}$, and $\Hilbert_2=\Hilbert_{S_i^c}$. By assumption \eqref{mostMITEcondition}, the probability that, for a particular $i$, the total error $\tilde\varepsilon + d_i/\sqrt{d_{\mc}}$ is greater than $\varepsilon$ is at most
\be
4 \exp\Bigl(-\frac{d_{\mc}\varepsilon^2}{72\pi^3}\Bigr)\,.
\ee 
The probability that this happens for any $i=1,\ldots,r$ is at most $r$ times this quantity, which completes the proof of \eqref{mostMITE}.

It may be surprising that the subsystems $S_i$ do not have to be very small for canonical typicality to hold but can, in fact, take up almost half of the whole system. For example, suppose that the system consists of a lattice of $N\gg 1$ spins, so $\dim \Hilbert =2^N$; suppose further that the energy shell arises from partitioning the energy axis into $10^{60} = 2^{200}$ intervals, so that, roughly, $d_{\mc} = 2^{N-200}$. If a subsystem $S_i$ consists of some subset of the $N$ spins comprising $49\%$ of the lattice sites, then
\be
d_i=2^{0.49N} \ll 2^{0.5N-100} = \sqrt{d_{\mc}}\,,
\ee 
so \eqref{mostMITEcondition} is satisfied. In fact, if we consider $r=10$ such subsystems of equal size and $\varepsilon = 10^{-12} = 2^{-40}$, then \eqref{mostMITEcondition} is satisfied for $N>14100$. 

(This leads to the question how large $r$ can be in \eqref{MITElistdef}. Continuing with the numbers just mentioned but dropping the assumption $r=10$, we obtain from \eqref{mostMITE} that $u_{\mc}(\MITE_{S_1,\ldots,S_r})\geq 1-10^{-30}$ for $r<\exp(2^{N-292}-71)$, which for large $N$ allows us to include \emph{all} sets of lattice sites comprising no more than $49\%$ of all sites. However, the definition of MITE in Section~\ref{sec:MITE1} required the appropriate behavior only for spatial regions of diameter $\leq \ell_0$, and as mentioned in Remark~\ref{rem:subsub}, a rather small number $r$ of regions of near-half volume, say $r=8$ for a system in a cube-shaped volume, will contain all regions of small diameter.)

So, for a subsystem $S$ comprising $49\%$ of the lattice sites, we have that for most $\psi\in\SSS(\Hilbert_{\mc})$,
\be
\hat\rho^\psi_S \approx \hat\rho^{\mc}_S \not\approx \hat\rho^{(\beta)}_S\,.
\ee
That is, while the density matrix obtained from $\psi$ is close to that from the micro-canonical ensemble, the latter is not necessarily 
close to that obtained from the canonical ensemble 
for any $\beta$. In fact, the canonical density matrix arises from $\hat\rho^{\mc}$ for \emph{small} subsystems $S$ (if the interaction between $S$ and $S^c$ is not too large), and $49\%$ of the lattice sites is not small enough for this effect to occur. 

What about subsystems $S$ greater than half of the whole system (say, comprising $51\%$ of the lattice sites, so $S^c$ is still a macroscopic system)? Is $\hat\rho^\psi_S\approx \hat\rho^{\mc}_S$ still true of most $\psi\in\SSS(\Hilbert_{\mc})$? The condition \eqref{mostMITEcondition} is then not fulfilled, but that may have been a merely sufficient condition. So here is an argument showing that canonical typicality will usually fail for subsystems greater than half of the whole. Suppose $\Hilbert_{\mc} = \Hilbert =\Hilbert_S\otimes \Hilbert_{S^c}$ with $d=\dim \Hilbert=2^N$ and $d_S=\dim\Hilbert_S = 2^{0.51N}$. For typical $\psi\in\SSS(\Hilbert)$, by canonical typicality $\hat\rho^{\psi}_{S^c}\approx \hat\rho^{\mc}_{S^c} = d_{S^c}^{-1} \hat I_{S^c} = (d_S/d) \hat I_{S^c}$. By the Schmidt decomposition, $\hat\rho^\psi_S$ has the same nonzero eigenvalues as $\hat\rho^\psi_{S^c}$, which are $d/d_S=2^{0.49N}$ nonzero eigenvalues of size $d_S/d=2^{-0.49N}$, whereas $\hat\rho^{\mc}_S = d_S^{-1} \hat I_S$ has $d_S=2^{0.51N}$ nonzero eigenvalues of size $2^{-0.51N}$, so $\hat\rho^\psi_S\not\approx \hat\rho^{\mc}_S$.

Realistic values for $d_{\mc}$ are
\be\label{dmcvalues}
\text{between }d_{\mc} = 10^{N/10} \quad \text{and} \quad
d_{\mc} = 10^{30 N}
\ee
(and thus something like $d_{\mc}=10^{10^{20}}$ or larger). Here are simple reasonings leading to these value. First, consider $N$ spins, so $\dim \Hilbert=2^N=10^{0.3 \, N}$, and suppose $d_{\mc}=(\dim\Hilbert)^{1/2}$. Second, for a single particle of mass $m$ in 1 dimension enclosed in a box of length $L$, the energy levels are $E_n = \hbar^2 \pi^2 n^2/ 2m L^2$. Thus, the energy levels of $N$ non-interacting particles in a 3-dimensional cubic box of side length $L$ are $(\hbar^2 \pi^2/2m L^2)\sum_{a=1}^3 \sum_{i=1}^N n_{ia}^2$, and the number $n$ of levels up to energy $E$ is approximately equal to the volume of the part with positive coordinates of a $3N$-dimensional ball of radius $R=L\sqrt{2mE}/\hbar\pi$ around the origin; this volume is $\approx 2^{-3N}\pi^{3N/2}R^{3N}/(3N/2)!\approx (e\pi R^2/6N)^{3N/2}$. For $E=\tfrac{3}{2}NkT$ with $T$ the temperature and $k$ Boltzmann's constant, we obtain $n\approx (3eL^2mkT/2\pi\hbar^2)^{3N/2}$. Thus, the number of levels in an energy interval of size $\Delta E = \tfrac{3}{2}Nk\Delta T$ is $n_{\Delta T}\approx (3N/2)(3eL^2mk/2\pi\hbar^2)^{3N/2} \, T^{3N/2-1}\, \Delta T$. For $\Delta T = 10^{-2}\,\mathrm{K}$, $T=300\,\mathrm{K}$, $L=1\,\mathrm{m}$, and $m=5\times 10^{-26}\,\mathrm{kg}$ (the mass of a nitrogen molecule), we obtain that $n_{\Delta T}\approx N 10^{33.6 N - 4.3}$; for a cubic meter of air, $N=2\times 10^{25}$, so $n_{\Delta T} \approx 10^{10^{25}}$.

\subsection{MITE for Abstract Subsystems}

A natural mathematical generalization that is often interesting to consider is based on dropping the idea that $S$ corresponds to a region in 3-space and regarding $S$ as an abstract subsystem defined by any splitting of Hilbert space into a tensor product,
\be
\Hilbert_{\mc} \subseteq \Hilbert_S \otimes \Hilbert_{S^c}\,,
\ee
where $S$ and $S^c$ can be thought of as just labels for the two factor spaces. For example, $S$ may comprise the spin degrees of freedom and $S^c$ the position degrees of freedom, or $S$ may comprise the oxygen atoms and $S^c$ all other atoms in the system. Then, canonical typicality as described in Remark~\ref{rem:can} or in \eqref{cantyp} still applies:
if $r$ is not too large and each $S_i$ is not too large ($\dim\Hilbert_{S_i} \ll \sqrt{\dim \Hilbert_{\mc}}$), then most $\psi\in\SSS(\Hilbert_{\mc})$ are ``in MITE relative to $S_1,\ldots,S_r$,'' i.e., lie in the set \eqref{MITElistdef}.

\subsection{MITE for Most Abstract Subsystems}

One can also consider the set $\MITE_{\text{most}}$ comprising those $\psi\in\SSS(\Hilbert_{\mc})$ for which $\hat{\rho}_S^\psi \approx \hat\rho^{\mc}_S$ holds for \emph{most} abstract subsystems $S$ with $\dim \Hilbert_S \leq d_0$. That is, instead of demanding $\psi\in\MITE_{S_i}$ for $r$ \emph{particular} subsystems $S_i$, we demand that $\psi\in\MITE_S$ for \emph{most} $S$. The key fact is that if $d_0\ll \sqrt{\dim\Hilbert_{\mc}}$, then 
\be
\text{most }\psi\in\SSS(\Hilbert_{\mc})\text{ lie in }\MITE_{\text{most}}\,. 
\ee
This claim follows from canonical typicality. Indeed, let $\sS$ be the set of all abstract subsystems $S$ of dimension $\leq d_0$, and let $\mu$ be the normalized uniform distribution over $\sS$. Since for every $S\in\sS$, most $\psi$ lie in $\MITE_S$ by canonical typicality, it follows from Fubini's theorem that under the product measure $\mu\times u_{\mc}$ on $\sS\times\SSS(\Hilbert_{\mc})$, the set of pairs $(S,\psi)$ such that $\psi\in\MITE_S$ has measure close to 1, and further that, for most $\psi$, $\mu\{S\in\sS: \psi\in\MITE_S\}\approx 1$. 

On the other hand, a pure state $\psi\in\SSS(\Hilbert_{\mc})$ cannot simultaneously lie in $\MITE_{S}$ for \emph{every} abstract subsystem $S$ of dimension $\leq d_0$. 
Put differently, for any given $\psi$ we can construct a subsystem $S$ for which $\psi$ is atypical. The simplest way of seeing this is to start with any given subsystem $S'$; then to find a $\psi'\in \Hilbert_{\mc}$ that is atypical for $S'$ in that $\hat\rho^{\psi'}_{S'}$ is far from $\hat\rho^{\mc}_{S'}$, for example $\psi'\approx \varphi\otimes \chi$ with $\varphi\in\Hilbert_{S'}$ and $\chi\in\Hilbert_{(S')^c}$; then to find a unitary operator $\hat U$ on $\Hilbert_{\mc}$ so that $\hat U \psi' = \psi$; and finally to define $S$ by applying $\hat U$ to $S'$. Another counterexample is described in \cite{Lych}.

\subsection{Remarks}

\begin{enumerate}
\setcounter{enumi}{\theremarks}
\item \emph{Superpositions of contributions from different energy shells.} Of course, some vectors in $\Hilbert$ have significant contributions from $\Hilbert_{\mc}$ for several macroscopically different energies $E$. In this paper, we focus on vectors in a single energy shell, as the implications for such superpositions are straightforward.

\item \emph{Local thermal equilibrium.} One often considers situations of local thermal equilibrium, in which for example the temperature is not constant throughout the volume occupied by the system, but varies slowly in space and time, and small regions can be regarded as being in thermal equilibrium. For such situations, there are then two different notions of local thermal equilibrium, corresponding to MITE and MATE.

\item \emph{Macro values are almost constant in the micro-canonical ensemble, micro values are random.} For every observable $\hat O$, $\hat\rho^{\mc}$ defines a probability distribution over its eigenvalues, the micro-canonical distribution; viz., the probability of eigenvalue $\alpha$ being
\be\label{probalphamc}
p^{\mc}(\alpha) = \tr(\hat\rho^{\mc}\hat{P}_\alpha)
\ee
with $\hat{P}_\alpha$ the eigenprojection for eigenvalue $\alpha$ of $\hat O$. For a (coarse-grained) macro observable $\hat M$, this distribution is almost constant, i.e., one value $\alpha_0$ has probability close to 1, and this value $\alpha_0$ is the thermal equilibrium value. For micro observables, in contrast, the distribution is not predominantly concentrated on a single value. For a macro observable $\hat O = \hat M$ again, when considering a pure state $\psi\in\SSS(\Hilbert_{\mc})$, the distribution over the eigenvalues, $p^\psi(\alpha) = \scp{\psi}{\hat{P}_\alpha|\psi}$, may be very different from $p^{\mc}(\alpha)$ for exceptional $\psi$, but for most $\psi$ it must again be predominantly concentrated on $\alpha_0$ because the micro-canonical distribution \eqref{probalphamc} equals the average of the $p^\psi$ over all $\psi\in\SSS(\Hilbert_{\mc})$.

\item \emph{Non-macroscopic systems.} While the thermodynamic ensembles $\hat\rho^{(\beta)}$ and $\hat\rho^{\mc}$ (or classically $\rho^{(\beta)}$ and $\rho^{\mc}$) can also be considered for a system that is non-macroscopic to begin with (say, that comprises only few particles), MATE (and its classical analog) are not defined for such a system because it does not have macro variables.\footnote{If we make an arbitrary choice of variables $\hat M_j$ instead, then these variables will usually not commute, not even approximately; and if they do commute, so that they define an orthogonal decomposition $\Hilbert=\oplus_\nu \Hilbert_\nu$, then the $\Hilbert_\nu$ will not feature the drastic differences in dimension (or the $\Gamma_\nu$ defined by an arbitrary choice of classical variables $M_j$ will not feature the drastic differences in volume) typical of macro-states, and there will usually not be a single macro-state that has 99.99\%\ of the size of the energy shell.} That is, the notion of MATE cannot be applied.

Concerning MITE, if the system is too small then canonical typicality will not apply (since canonical typicality requires that the ``bath,'' i.e., the complement of the subsystem, be large), and the set $\MITE$ may well be empty. However, MITE is well approximated in surprisingly small systems, such as 
for example the six atom, six site Bose--Hubbard chain studied experimentally in \cite{KTL16}.

\item{\it Other Measures of Typicality Than Micro-Canonical.}
We have mentioned that most $\psi\in\SSS(\Hilbert_{\mc})$ are in both MITE and MATE; put differently, MITE and MATE are typical properties relative to $u_{\mc}$, the uniform distribution over $\SSS(\Hilbert_{\mc})$. This distribution can be called the \emph{micro-canonical distribution} of wave functions, as it plays a role analogous to the micro-canonical distribution of phase points in classical mechanics. This brings us to the question whether, instead of starting out from $u_{\mc}$,
we could have started from another distribution. Is there a distribution of wave function analogous to the \emph{canonical} distribution of phase points in classical mechanics? And are MITE and MATE typical relative to that distribution? 

We conjecture that the answers are yes and yes. The natural candidate for the canonical distribution of wave functions is the measure known as $GAP(\hat\rho^{(\beta)})$ (``$G$aussian $A$djusted $P$rojected measure''). For any density operator $\hat\rho$ on $\Hilbert$, the measure $GAP(\hat\rho)$ \cite{JRW94,GLTZ06b,Rei08b,GLMTZ15}, called the ``Scrooge measure'' in \cite{JRW94}, is the most spread-out distribution on $\SSS(\Hilbert)$ that has density operator $\hat\rho$. For comparison, the least spread-out distribution would be concentrated on an eigenbasis of $\hat\rho$ with weights given by the eigenvalues of $\hat\rho$. When $\hat\rho$ is proportional to a projection, then $GAP(\hat\rho)$ is uniform over the sphere in the range of that projection; thus, $GAP(\hat\rho^{\mc})=u_{\mc}$. It turns out \cite{GLTZ06b,GLMTZ15} that for most $\psi\in\SSS(\Hilbert_{\mc})$, the conditional wave function of a small subsystem $S$ is approximately $GAP(\hat\rho_S^{\mc})$-distributed; in this way, this distribution is a quantum analog of the canonical distribution of phase points in classical mechanics, and one can say that $GAP(\hat\rho^{(\beta)})$ is the thermal equilibrium distribution of the wave function.

We conjecture that most $\psi$ relative to $GAP(\hat\rho^{(\beta)})$ have $\hat\rho^\psi_S \approx \hat\rho^{(\beta)}_S$ for small subsystems $S$. This parallel between the canonical and the micro-canonical distribution of wave functions would be some kind of equivalence of ensembles. However, we note that a $u_{\mc}$-typical $\psi^{\mc}$ looks quite different from a $GAP(\hat\rho^{(\beta)})$-typical $\psi^{(\beta)}$: While $\psi^{\mc}$ lies in $\Hilbert_{\mc}$, $\psi^{(\beta)}$ does not; while the coefficients of $\psi^{\mc}$ in the energy eigenbasis $\{\phi_\alpha\}$ are (with high probability) all of roughly equal magnitude (or zero), the coefficients $\scp{\phi_\alpha}{\psi^{(\beta)}}$ have rather different magnitudes, whose squares are roughly proportional to $e^{-\beta E_\alpha}$; as a consequence, more coefficients are nonzero, and more are significantly nonzero than for $\psi^{\mc}$. In fact, the energy uncertainty of $\psi^{\mc}$ is of order $1/\beta$ 
(independently of $N$ if we keep $\beta$ fixed), while the energy uncertainty of $\psi^{(\beta)}$ is proportional to $\sqrt{N}$; both are much smaller than the size $\Delta E$ of the energy window, which is proportional to $N$.
\end{enumerate}
\setcounter{remarks}{\theenumi}

\section{Exceptional Cases}
\label{sec:exceptional}

There are at least two exceptional situations in which a dominant macro-state $\Gamma_{\eq}$ or $\Hilbert_{\eq}$ does not exist. First, at a first-order phase transition, such as in the ferromagnetic Ising model in a vanishing external magnetic field, some $\Gamma_\nu$ (or $\Hilbert_{\nu}$) has the appropriate majority of spins up and some $\Gamma_{\nu'}$ (or $\Hilbert_{\nu'}$) has the appropriate majority of spins down, each having nearly 50\%\ of the volume of $\Gamma_{\mc}$ (of the dimension of $\Hilbert_{\mc}$) for a suitable energy interval. 

Second, if the size of the system is exorbitant, say its volume is greater than $10^{10^{10}}$ cubic meters\footnote{Of course, already at much smaller sizes than that, another phenomenon that we are neglecting in this paper becomes very relevant: gravity. It was for this reason that Onsager wrote \cite{Ons}: ``[T]he common concept of a homogeneous volume phase implies dimensions that are large compared to the molecules and small compared to the moon.''} (which is about $10^{10^{10}}$ times the volume of the known universe, which is $10^{80}$ cubic meters), while we keep the size of the cells $\Lambda_j$ small on the macro scale, then the number of cells will be correspondingly large, and it is to be expected by chance alone that a uniformly-randomly selected phase point in $\Gamma_{\mc}$ will possess a cell $\Lambda_j$ somewhere in which a macroscopic observable $M_j$ deviates significantly from its average value. As a consequence, the set where \emph{every} $M_j$ assumes its average value will not have most of the volume. Likewise, for a randomly selected $\psi\in\SSS(\Hilbert_{\mc})$ in such an exorbitantly large system, the joint probability distribution that $\psi$ defines over the eigenvalues $\nu_j$ of the macroscopic observables $\hat M_j$ will not be overwhelmingly concentrated on a single $(\nu_1,\ldots,\nu_K)$. 

To obtain the estimate that $10^{10^{10}}$ cubic meters is the relevant volume (say, in the classical case), we subdivide the volume into $m$ cells of (say) cubic millimeter size, consider the volume filled with air at room conditions, which has $n\approx 2.5\times 10^{16}$ particles (i.e., $N_2$ molecules) per cubic millimeter, and ask whether the number of particles in any cell will be less than $0.999n$ or more than $1.001n$. Since for a random phase point, the particles will be essentially uniformly distributed over the volume, the number $N_i$ of particles in cell $i$ has a binomial distribution with parameters $nm$ and $m^{-1}$, which for large $n$ and $m$ is approximately Gaussian with mean $n$ and variance $n$. The probability that $N_i<0.999n$ or $N_i>1.001n$ is of order $e^{-(0.001n)^2/2n}=e^{-n/2\times 10^6}$, so for an appreciable probability that this happens for any cell anywhere, we need that $m \gtrsim e^{n/2\times 10^6} \approx 10^{10^{10}}$.

This effect, that  for exorbitantly large systems none of the $\Gamma_\nu$ or $\Hilbert_\nu$ is dominant, can be problematical when we want to take the thermodynamic limit and let the volume tend to infinity. It can easily be dealt with, either by increasing the cell size and the tolerances $\Delta M_j$ as we take the limit, or by defining $\Gamma_{\eq}$ differently as the set of those $X\in\Gamma_{\mc}$ at which \emph{most}, but not \emph{all}, macro observables $M_j$ assume their thermal equilibrium values (and $\Hilbert_{\eq}$ as the subspace of $\Hilbert_{\mc}$ on which \emph{most}, but not \emph{all}, macro observables $\hat M_j$ assume their thermal equilibrium values). 

This effect also entails that the notion of MATE becomes meaningless for exorbitantly large systems (unless we increase cell size and tolerances or redefine $\Hilbert_{\eq}$), while MITE remains unaffected by this situation. Indeed, by virtue of the theorem of Popescu et al.~\cite{PSW05,PSW06} about canonical typicality (see Section~\ref{sec:MITE} below), the probability that for a (say, cubic millimeter sized) 3-cell $\Lambda_i$, $\|\hat\rho^\psi_{\Lambda_i} - \hat\rho^{\mc}_{\Lambda_i}\| > \varepsilon$ is, for fixed small $\varepsilon>0$, of order $\exp(-\varepsilon^2 d_{\mc})$ as $d_{\mc}=\dim\Hilbert_{\mc} \to\infty$. Thus, if we consider $m$ cells, the probability that any of them will be subject to a deviation $\|\hat\rho^\psi_{\Lambda_i} - \hat\rho^{\mc}_{\Lambda_i}\| > \varepsilon$ is at most $m \exp(-\varepsilon^2 d_{\mc})$, and since $d_{\mc}$ is of order $m^\lambda e^{\kappa m}$ with $\kappa,\lambda>0$ as we keep the cell size while increasing the number of cells (and thus the system size), that probability gets small as $m\to \infty$. Thus, as $m\to\infty$ it has probability close to 1 for random $\psi\in\SSS(\Hilbert_{\mc})$ that \emph{all} cells will simultaneously be close to thermal equilibrium in the sense $\hat\rho^\psi_{\Lambda_i} \approx \hat\rho^{\mc}_{\Lambda_i}$. 

So this is another difference between MITE and MATE: MATE becomes meaningless for exorbitantly large systems (unless we change the cell size and tolerances, or the definition of $\Hilbert_{\eq}$) and MITE does not. As a consequence, since MATE but not MITE exists in classical mechanics for pure states, it is also a difference between the quantum and the classical case: 
for an exorbitantly large system, the notion of thermal equilibrium for pure states becomes problematical in classical mechanics but not (in the sense of MITE) in quantum mechanics.

\section{Other Proposed Definitions of Thermal Equilibrium}
\label{sec:otherdefs}

\subsection{Tasaki's Version of MATE} 
\label{sec:TMATE}

Tasaki \cite{Tas15,Tas15b} noted that there can be substantial practical difficulty about finding, for a specific example of a physical system, a realistic orthogonal decomposition \eqref{decomp} and proving that one of the macro-spaces $\Hilbert_\nu$ in $\Hilbert_{\mc}$ has $>99\%$ of the dimensions. He suggested the following alternative definition (see \cite{DRMN06,Sug07} for earlier work in this direction), which is not strictly but approximately equivalent to MATE and which we call TMATE: For any collection $\hat M_1,\ldots, \hat M_K$ of self-adjoint operators (thought of as representing macro observables but not necessarily commuting), we say that a system with state $\hat\rho$ is in TMATE if and only if
\be\label{TMATEdef}
\tr(\hat\rho \hat P_j) > 1-\delta \quad \forall j=1,\ldots,K,
\ee
where
\be
\hat P_j = 1_{[V_j-\Delta M_j,V_j+\Delta M_j]}(\hat M_j)\,,
\ee
\be
V_j = \tr(\hat\rho^{\mc}\, \hat M_j)
\ee
is the thermal equilibrium value of $\hat M_j$, and $1_A$ denotes the characteristic function (indicator function) of the set $A$. Note that $\hat P_j$ is the projection to the subspace spanned by the eigenspaces of $\hat M_j$ with eigenvalues within $V_j\pm \Delta M_j$; thus, 
$\tr(\hat\rho \hat P_j)$ is the probability of finding, in a quantum measurement of $\hat M_j$ on a system in state $\hat\rho$, a value within $V_j \pm \Delta M_j$. In particular, the set of pure states in TMATE is given by
\be\label{TMATEpuredef}
\TMATE = \bigcap_{j=1}^K \Bigl\{ \psi\in\SSS(\Hilbert_{\mc}): \scp{\psi}{\hat P_j|\psi}>1-\delta \Bigr\}\,.
\ee
If, for each $j$, the range of $\hat P_j$ has almost full dimension (as did $\Hilbert_{\eq}$ in our previous conderations, and as it should be the case for a macro observable $\hat M_j$ and a macroscopic tolerance $\Delta M_j$), then most $\psi\in\SSS(\Hilbert_{\mc})$ lie in \eqref{TMATEpuredef}. That is, quantitatively, if the dimension of the range of $\hat P_j$ is greater than $(1-\varepsilon/\delta) \dim \Hilbert_{\mc}$ for each $j$, then $u_{\mc}(\TMATE) > 1-K\varepsilon/\delta$, which is close to 1 if $\varepsilon \ll \delta/K$.

The basic point of TMATE is that the procedures involved in the choice of the subspace $\Hilbert_{\eq}$, such as rounding off the $\hat M_j$ to make them commute, are not crucial for obtaining a workable version of MATE, so that TMATE is simpler than MATE as defined in \eqref{MATEdef} from the perspective of practical computation, while keeping the essence of the concept of MATE.

\subsection{Von Neumann's Proposed Definition}

Von Neumann \cite{vN29} proposed a further definition of thermal equilibrium, inequivalent to MITE and MATE, that is also based on an orthogonal decomposition $\Hilbert_{\mc}=\oplus_\nu \Hilbert_\nu$ into the simultaneous eigenspaces of a commuting family $\{\hat M_1,\ldots, \hat M_K\}$ of macro observables. According to this definition, a system with pure state $\psi\in\SSS(\Hilbert_{\mc})$ is in thermal equilibrium if and only if
\be\label{normal}
\|\hat P_\nu\psi\|^2 \approx \frac{\dim \Hilbert_\nu}{\dim \Hilbert_{\mc}}
\quad \text{for all }\nu,
\ee
where $\hat P_{\nu}$ is the projection to $\Hilbert_\nu$ and $a\approx b$ can be taken to mean (say) $0.99<a/b<1.01$.
See \cite{GLMTZ10,GLTZ10} for discussion of the property \eqref{normal}, called there \emph{normality}. Suppose that among the $\Hilbert_\nu$ there is a dominant subspace $\Hilbert_{\eq}$. Then von Neumann's equilibrium states all lie in $\MATE$, and their macroscopic behavior is practically indistinguishable from other states in $\MATE$, which is why $\MATE$ then seems like the more natural definition. 

Von Neumann considered only the case in which there is no dominant $\Hilbert_\nu$, which occurs if one takes the inaccuracies $\Delta M_j$ in the coarse-graining involved in the construction of the macro observables $\hat M_j$ smaller than the typical size of fluctuations in thermal equilibrium. That is, a smaller choice of $\Delta M_j$ corresponds to a finer partition of $\Gamma$ into $\Gamma_\nu$ or of $\Hilbert$ into $\Hilbert_\nu$, and for sufficiently small $\Delta M_j$, none of the $\Gamma_\nu$ or $\Hilbert_\nu$ will have 99\%\ of the size of the energy shell. According to the estimate
\be
\varepsilon=e^{-mN\Delta M_j^2}
\ee
of Section~\ref{sec:MATE}, this may happen if
\be
\Delta M_j \lesssim \frac{1}{\sqrt{mN}}\,,
\ee
i.e., if
\be
\text{relative error} = \frac{\Delta M_j}{\nu_j^{\eq}} = m\Delta M_j \lesssim \sqrt{\frac{m}{N}}
\ee
with $\nu_j^{\eq}$ the eigenvalue of $\hat M_j$ on $\Hilbert_{\eq}$; this means a relative error of $3\times 10^{-6}$ or less for $m=10^9$ (number of 3-cells) and $N=10^{20}$ (number of particles). That a macroscopic measurement could determine the number of particles in a given cubic millimeter of a macroscopic system (or the amount of energy, or charge, or magnetization in that volume) with an accuracy of 6 digits seems not realistically feasible, so the assumption of such a small $\Delta M_j$ is perhaps overly stretching the idea of ``macroscopic.''

This leads us to another difference between MITE and MATE: If the $\Delta M_j$ are chosen so small (as von Neumann had in mind) that none of the macro-spaces $\Hilbert_\nu$ becomes dominant, 
then MATE cannot be applied any more, while MITE still can. This situation is parallel to that discussed in Section~\ref{sec:exceptional} above.

It seems that Reimann's \cite{Rei15b} recent approach using a typical observable $\hat A$ is closely related to von Neumann's if we consider an orthogonal decomposition $\oplus_\nu \Hilbert_\nu$ that arises not from commuting macro observables but instead as the eigenspaces of the single observable $\hat A$.

\section{Conclusions}
\label{sec:conclusions}

Arguably, MATE is the more immediate concept of thermal equilibrium. After all, thermal equilibrium is a notion of thermodynamics, and its meaning there is that the macro appearance of the system is stationary, and that temperature and chemical potential are spatially uniform (understood in terms of the spatial distribution of energy). This meaning corresponds to MATE, not to MITE.

Moreover, the notion of thermal equilibrium is not exclusive to quantum mechanics, as thermal equilibrium is equally possible in classical mechanics, and in fact the concept originated in classical mechanics; so the definition of thermal equilibrium may be expected to be a general one that applies to both classical and quantum mechanics. This would be so for MATE but not for MITE (which does not exist in classical mechanics for pure states).

On the other hand, since MITE is the stronger property, and since it is usually true that macroscopic quantum systems approach MITE (MBL systems being an exception), it is natural to consider MITE, and it would seem artificial to not regard it 
as a new kind of thermal equilibrium property emerging from quantum entanglement.

For MBL systems, most energy eigenstates $\phi_\alpha$ have a short range of entanglement. Usually, some $\phi_\alpha$'s of MBL systems are not in MATE (so states with significant contribution from them will not thermalize), and in fact some $\phi_\alpha$'s are even approximately orthogonal to $\Hilbert_{\eq}$. 
Since the $\phi_\alpha$'s are more or less product states of eigenstates of local (cell) energy, they lack the long-range entanglement relevant to MITE, and thus almost all fail to satisfy MITE. Yet, considering, instead of energy eigenstates, typical wave functions $\psi$ from an energy shell, they do feature long-range entanglement and thus are in MITE, and a fortiori in MATE.

We note finally that while our analysis has focused exclusively on macroscopic systems, there is strong numerical and even experimental evidence~\cite{JS,KTL16} that MITE can be a very good approximation for surprisingly small quantum systems of just a few spins or a few atoms, even in pure states.
For such systems MATE is not defined at all in either classical or quantum mechanics. It is also not clear whether (and if so how) the concepts of ``thermodynamics,"  of Boltzmann entropy, and of the second law can be applied to such an isolated microscopic quantum system.

\bigskip

\noindent\textit{Acknowledgments.} 
D.~A.~Huse is the Addie and Harold Broitman Member at IAS. 
J.~L.~Lebowitz was supported in part by the National Science Foundation [grant  DMR1104501] and AFOSR [grant F49620-01-0154].


\begin{thebibliography}{18}

\bibitem{And58} P. W. Anderson:
	Absence of Diffusion in Certain Random Lattices.
	\textit{Physical Review} \textbf{109}: 1492--1505 (1958)

\bibitem{BAA06a} D. M. Basko, I. L. Aleiner, and B. L. Altshuler:
	Metal--insulator transition in a weakly interacting many-electron system 
	with localized single-particle states.
	Annals of Physics \textbf{321}: 1126--1205 (2006)

\bibitem{BAA06b} D. M. Basko, I. L. Aleiner, and B. L. Altshuler:
	On the problem of many-body localization.
	Pages 50--69 in A. L. Ivanov and S. G. Tikhodeev (editors):
	\textit{Problems of Condensed Matter Physics}, Oxford University Press (2008)
	\url{http://arxiv.org/abs/cond-mat/0602510}

\bibitem{BN13} B. Bauer and C. Nayak:
	Area laws in a many-body localized state and its implications for topological order.
	\textit{Journal of Statistical Mechanics} P09005 (2013) 
	\url{http://arxiv.org/abs/1306.5753}

\bibitem{B96} L. Boltzmann:
                            \emph{Vorlesungen \"{u}ber Gastheorie},
                            2 volumes. Leipzig: Barth (1896, 1898).
                            English translation by S. G. Brush: \emph{Lectures on Gas
                            Theory.} Cambridge University Press (1964)

\bibitem{Borel} E. Borel:
	\textit{Probabilities and life.}
	London: Dover (1962)

\bibitem{Ca} H. B. Callen:
	Thermodynamics and an Introduction to Thermostatistics.
	New York: Wiley (1985)

\bibitem{DRMN06} W. De Roeck, C. Maes, and K. Neto\v{c}n\'y:  
	Quantum Macrostates, Equivalence of Ensembles and an $H$-theorem.
	\textit{Journal of Mathematical Physics} \textbf{47}: 073303 (2006)
	\url{http://arxiv.org/abs/math-ph/0601027}

\bibitem{Deu91} J. M. Deutsch:
	Quantum statistical mechanics in a closed system.
	\textit{Physical Review A} \textbf{43}: 2046--2049 (1991)

\bibitem{Ein14} A. Einstein: 
	Beitr\"age zur Quantentheorie.
	\textit{Deutsche Physikalische Gesellschaft. Verhandlungen} 
	\textbf{16}: 820--828 (1914). 
	English translation in 
	\textit{The Collected Papers of Albert Einstein}, Vol.~6, pp.~20--26.
	Princeton: University Press (1996) 

\bibitem{GMM04} J. Gemmer, G. Mahler, and M. Michel:
    \textit{Quantum Thermodynamics:
      Emergence of Thermodynamic Behavior within Composite Quantum Systems}.
      Lecture Notes in Physics \textbf{657}. Berlin: Springer (2004)

\bibitem{GE15} C. Gogolin and J. Eisert:
	Equilibration, thermalisation, and the emergence of statistical mechanics 
	in closed quantum systems.
	\textit{Reports on Progress in Physics} \textbf{79}: 056001 (2016)
	\url{http://arxiv.org/abs/1503.07538}

\bibitem{Gol98} S. Goldstein: 
	Quantum Theory Without Observers.
	\textit{Physics Today}, Part One: March 1998, 42--46. 
	Part Two: April 1998, 38--42.

\bibitem{Gol99} S. Goldstein:
	Boltzmann's approach to statistical mechanics.
	Pages 39--54 in J. Bricmont, D. D\"urr, M. C. Galavotti, G. C. Ghirardi, 
	F. Petruccione, and N. Zangh\`\i\ (ed.s), 
	\textit{Chance in Physics: Foundations and Perspectives}, 
	\textit{Lecture Notes in Physics} {\bf 574}. Berlin: Springer-Verlag (2001)
	\url{http://arxiv.org/abs/cond-mat/0105242}

\bibitem{GHT13b} S. Goldstein, T. Hara, H. Tasaki:
	Time scales in the approach to equilibrium of macroscopic quantum systems.
	\textit{Physical Review Letters}  \textbf{111}: 140401 (2013)
	\url{http://arxiv.org/abs/1307.0572}

\bibitem{GHT14a} S. Goldstein, T. Hara, H. Tasaki:
	Extremely quick thermalization in a macroscopic quantum system for 
	a typical nonequilibrium subspace.
	\textit{New Journal of Physics} \textbf{17}: 045002 (2015) 
	\url{http://arxiv.org/abs/1402.0324}

\bibitem{GHT14b} S. Goldstein, T. Hara, H. Tasaki:
	The approach to equilibrium in a macroscopic quantum system for 
	a typical nonequilibrium subspace.
	\url{http://arxiv.org/abs/1402.3380}

\bibitem{GHLT15a} S. Goldstein, D. A. Huse, J. L. Lebowitz, and R. Tumulka:
	Thermal Equilibrium of a Macroscopic Quantum System in a Pure State.
	\textit{Physical Review Letters} \textbf{115}: 100402 (2015)
	\url{http://arxiv.org/abs/1506.07494}

\bibitem{GL} S. Goldstein and J. L. Lebowitz:
	On the (Boltzmann) Entropy of Nonequilibrium Systems.
	\textit{Physica D} \textbf{193}: 53--66 (2004)
	\url{http://arxiv.org/abs/cond-mat/0304251}

\bibitem{GLMTZ10} S. Goldstein, J. L.~Lebowitz, C. Mastrodonato,
	R. Tumulka, and N. Zangh\`\i: 
	Normal Typicality and von Neumann's Quantum Ergodic Theorem.
	\textit{Proceedings of the Royal Society A} \textbf{466}: 3203--3224 (2010)
	\url{http://arxiv.org/abs/0907.0108}

\bibitem{GLMTZ09b} S. Goldstein, J. L. Lebowitz, C. Mastrodonato, R. Tumulka, and N. Zangh\`\i:
	Approach to Thermal Equilibrium of Macroscopic Quantum Systems.
	\textit{Physical Review E} \textbf{81}: 011109 (2010)
	\url{http://arxiv.org/abs/0911.1724}

\bibitem{GLMTZ15} S. Goldstein, J. L. Lebowitz, C. Mastrodonato, R. Tumulka, and N. Zangh\`\i:
	Universal Probability Distribution for the Wave Function of a Quantum
	System Entangled with Its Environment.
	{\it Communications in Mathematical Physics} {\bf 342}: 965--988 (2016)
	\url{http://arxiv.org/abs/1104.5482}

\bibitem{GLTZ06} S. Goldstein, J.L. Lebowitz, R. Tumulka, and N. Zangh\`\i:
	Canonical Typicality.
	\textit{Physical Review Letters} \textbf{96}: 050403 (2006)
	\url{http://arxiv.org/abs/cond-mat/0511091}

\bibitem{GLTZ06b} S. Goldstein, J.L. Lebowitz, R. Tumulka, and N. Zangh\`\i:
	On the Distribution of the Wave Function for Systems in Thermal Equilibrium.
	\textit{Journal of Statistical Physics} \textbf{125}: 1193--1221 (2006)
	\url{http://arxiv.org/abs/quant-ph/0309021}

\bibitem{GLTZ10} S. Goldstein, J. L. Lebowitz, R. Tumulka, and N. Zangh\`\i:
	Long-Time Behavior of Macroscopic Quantum Systems.
	\textit{European Physical Journal H} \textbf{35}: 173--200 (2010)
	\url{http://arxiv.org/abs/1003.2129}

\bibitem{Gri} R. Griffiths: Statistical Irreversibility: Classical and Quantum.
	Pages 147--159 in J. J. Halliwell, J. P\'erez-Mercader, and W. H. Zurek (editors):
	\textit{Physical Origin of Time Asymmetry}.
	Cambridge University Press (1994)

\bibitem{hno} D. A. Huse, R. Nandkishore and V. Oganesyan:
	Phenomenology of fully many-body-localized systems.
	\textit{Physical Review B} \textbf{90}: 174202 (2014)
	\url{http://arxiv.org/abs/1408.4297}

\bibitem{Imbrie} J.Z. Imbrie:
	On Many-Body Localization for Quantum Spin Chains.
	\textit{Journal of Statistical Physics} \textbf{163}: 998--1048 (2016)
	\url{http://arxiv.org/abs/1403.7837}

\bibitem{JS} R. V. Jensen and R. Shankar: 
	Statistical Behavior in Deterministic Quantum Systems with Few Degrees of Freedom.
	\textit{Physical Review Letters} \textbf{54}: 1879--1882 (1985)

\bibitem{JRW94} R. Jozsa, D. Robb, and W.K. Wootters:
	Lower bound for accessible information in quantum mechanics.
	\textit{Physical Review A} \textbf{49}: 668--677 (1994)

\bibitem{KTL16} A. M. Kaufman, M. E. Tai, A. Lukin, M. Rispoli, R. Schittko, P. M. Preiss, and M. Greiner:
	Quantum thermalization through entanglement in an isolated many-body system.
	{\it Science} {\bf 353}: 794--800 (2016)
	\url{http://arxiv.org/abs/1603.04409}

\bibitem{KIH14} H. Kim, T. N. Ikeda, and D. A. Huse:
	Testing whether all eigenstates obey the Eigenstate Thermalization Hypothesis.
	\textit{Physical Review E} \textbf{90}: 052105 (2014)
	\url{http://arxiv.org/abs/1408.0535}

\bibitem{Lan} O. E. Lanford: 
	Entropy and Equilibrium States in Classical Statistical Mechanics. 
	Pages 1--113 in A. Lenard (ed.), \textit{Lecture Notes in Physics} \textbf{2}. 
	Berlin: Springer-Verlag (1973)

\bibitem{L07} J. L. Lebowitz:
	From Time-symmetric Microscopic Dynamics to Time-asymmetric 
	Macroscopic Behavior: An Overview.
	Pages 63--88 in G.~Gallavotti , W.~L.~Reiter, J.~Yngvason (editors): 
	\textit{Boltzmann's Legacy}.
	Z\"urich: European Mathematical Society (2008)
	\url{http://arxiv.org/abs/0709.0724}

\bibitem{LM03} J. L. Lebowitz and C. Maes: 
	Entropy---A Dialogue. 
	Pages 269--273 in A.~Greven, G.~Keller, G.~Warnecke (editors): 
	\textit{On Entropy.} 
	Princeton: University Press (2003)
	
\bibitem{LPSW08}
	N. Linden, S. Popescu, A. J. Short, and A. Winter:
	Quantum mechanical evolution towards thermal equilibrium. 
	\textit{Physical Review E} \textbf{79}: 061103 (2009)
	\url{http://arxiv.org/abs/0812.2385}

\bibitem{Lych} O. Lychkovskiy:
	Dependence of decoherence-assisted classicality on the way a system is 
	partitioned into subsystems.
	\textit{Physical Review A} \textbf{87}: 022112 (2013)
	\url{http://arxiv.org/abs/1210.4124}

\bibitem{NH15} R. Nandkishore and D. A. Huse:
	Many body localization and thermalization in quantum statistical mechanics.
	\textit{Annual Review of Condensed Matter Physics} {\bf 6}: 15--38 (2015)
	\url{http://arxiv.org/abs/1404.0686}

\bibitem{OH07} V. Oganesyan and D. A. Huse:
	Localization of interacting fermions at high temperature.
	\textit{Physical Review B} \textbf{75}: 155111 (2007) 
	\url{http://arxiv.org/abs/cond-mat/0610854}

\bibitem{Y} Y.~Ogata:
	Approximating macroscopic observables in quantum spin systems with 
	commuting matrices.
	\textit{Journal of Functional Analysis} \textbf{264}: 2005--2033 (2013)
	\url{http://arxiv.org/abs/1111.5933}
    
\bibitem{Ons} L. Onsager: 
	Thermodynamics and Some Molecular Aspects of Biology.
	Pages 75--79 in G. C. Quarton et al. (editors):
	\textit{The Neurosciences: A Study Program.}
	New York: Rockefeller University Press (1967)

\bibitem{PH10} A. Pal and D. A. Huse:
	The many-body localization phase transition.
	\textit{Physical Review B} \textbf{82}: 174411 (2010)
	\url{http://arxiv.org/abs/1010.1992}

\bibitem{Pen04} R. Penrose:
	\textit{The Road to Reality.}
	London: Jonathan Cape (2004)

\bibitem{PSW05} S. Popescu, A. J. Short, and A. Winter:
	The foundations of statistical mechanics from entanglement:
	Individual states vs.\ averages.
	Preprint (2005) 
	\url{http://arxiv.org/abs/quant-ph/0511225}

\bibitem{PSW06} S. Popescu, A. J. Short, and A. Winter:
  Entanglement and the foundation of statistical mechanics.
  \textit{Nature Physics} \textbf{21(11)}: 754--758 (2006)

\bibitem{Rei07} P. Reimann:
  Typicality for Generalized Microcanonical Ensembles.
  \textit{Physical Review Letters} \textbf{99}: 160404 (2007)
  \url{http://arxiv.org/abs/0710.4214}

\bibitem{Rei08b} P. Reimann: 
	Typicality of pure states randomly sampled according to the 
	Gaussian adjusted projected measure. 
	\textit{Journal of Statistical Physics} \textbf{132(5)}: 921--935 (2008)
	\url{http://arxiv.org/abs/0805.3102}

\bibitem{Rei08} P. Reimann: 
	Foundation of Statistical Mechanics under Experimentally Realistic Conditions. 
	\textit{Physical Review Letters} \textbf{101}: 190403 (2008)
	\url{http://arxiv.org/abs/0810.3092}

\bibitem{Rei10} P. Reimann:
	Canonical thermalization.
	\textit{New Journal of Physics} \textbf{12}: 055027 (2010)
	\url{http://arxiv.org/abs/1005.5625}

\bibitem{Rei15} P. Reimann:
	Eigenstate thermalization: Deutsch's approach and beyond.
	\textit{New Journal of Physics} \textbf{17}: 055025 (2015)
	\url{http://arxiv.org/abs/1505.07627}

\bibitem{Rei15b} P. Reimann:
	Generalization of von Neumann's Approach to Thermalization.
	{\it Physical Review Letters} {\bf 115}: 010403 (2015)
	\url{http://arxiv.org/abs/1507.00262}

\bibitem{RDO08} M. Rigol, V. Dunjko, and M. Olshanii:
	Thermalization and its mechanism for generic isolated quantum systems.
	\textit{Nature} \textbf{452}: 854--858 (2008)
	\url{http://arxiv.org/abs/0708.1324}

\bibitem{RS12} M. Rigol and M. Srednicki: 
	Alternatives to Eigenstate Thermalization.
	\textit{Physical Review Letters} \textbf{108}: 110601 (2012)
	\url{http://arxiv.org/abs/1108.0928}

\bibitem{ros} V. Ros, M. M\"uller, and A. Scardicchio:
	Integrals of motion in the many-body localized phase.
	\textit{Nuclear Physics B} \textbf{891}: 420-465 (2015)
	\url{http://arxiv.org/abs/1406.2175}

\bibitem{Schr27} E. Schr\"odinger: 
	Energieaustausch nach der Wellenmechanik.
	\textit{Annalen der Physik} \textbf{83(15)}: 956--968 (1927).
	English translation by J. F. Shearer and W. M. Deans: 
	The Exchange of Energy according to Wave Mechanics, 
	pages 137--146 in E. Schr\"odinger: \textit{Collected Papers on Wave Mechanics}, 
	Providence, R.I.: AMS Chelsea (1982)

\bibitem{Schr52} E. Schr\"odinger:
	\textit{Statistical Thermodynamics.}
	Second Edition, Cambridge University Press (1952)

\bibitem{Sre94} M. Srednicki:
	Chaos and quantum thermalization.
	\textit{Physical Review E} \textbf{50}: 888--901 (1994)

\bibitem{Sre96} M. Srednicki:
	Thermal Fluctuations in Quantized Chaotic Systems.
	\textit{Journal of Physics A: Mathematical and General} \textbf{29}: L75--L79 (1996)
	\url{http://arxiv.org/abs/chao-dyn/9511001}

\bibitem{Sre99} M. Srednicki:
	The approach to thermal equilibrium in quantized chaotic systems.
	\textit{Journal of Physics A: Mathematical and General} \textbf{32}: 1163--1176 (1999)
	\url{http://arxiv.org/abs/cond-mat/9809360}

\bibitem{Sug07} A. Sugita:
	On the Basis of Quantum Statistical Mechanics.
	\textit{Nonlinear Phenomena in Complex Systems} \textbf{10}: 192--195 (2007)
	\url{http://arxiv.org/abs/cond-mat/0602625}

\bibitem{T98} H. Tasaki: 
	From Quantum Dynamics to the Canonical
	Distribution: General Picture and a Rigorous Example. 
	\textit{Physical Review Letters} \textbf{80}: 1373--1376 (1998)
	\url{http://arxiv.org/abs/cond-mat/9707253}

\bibitem{Tas10} H. Tasaki:
	The approach to thermal equilibrium and ``thermodynamic normality".
	\url{http://arxiv.org/abs/1003.5424}

\bibitem{Tas15} H. Tasaki, personal communication (2015)

\bibitem{Tas15b} H. Tasaki:
	Typicality of thermal equilibrium and thermalization in isolated macroscopic
	quantum systems.
	\textit{Journal of Statistical Physics} \textbf{163}: 937--997 (2016)
	\url{http://arxiv.org/abs/1507.06479}

\bibitem{Ull64} N. Ullah: 
	Invariance hypothesis and higher correlations of Hamiltonian matrix elements.
	{\it Nuclear Physics} {\bf 58}: 65 (1964)

\bibitem{vN29} J. von Neumann:
      Beweis des Ergodensatzes und des $H$-Theorems in der neuen Mechanik. 
      \textit{Zeitschrift f\"ur Physik} \textbf{57}: 30--70 (1929). 
      English translation in
      \textit{European Physical Journal H} \textbf{35}: 201--237 (2010)
	\url{http://arxiv.org/abs/1003.2133}

\end{thebibliography}
\end{document}